# On the Nature of Flow Curve and Categorization of Thixotropic Yield Stress Materials

Tulika Bhattacharyya, [a)] Alan R. Jacob, [b)] George Petekidis, * [c)] and Yogesh M. Joshi* [a)]

[a)] Department of Chemical Engineering, Indian Institute of Technology Kanpur, Kanpur, Uttar Pradesh 208016, India

[b)] Department of Chemical Engineering, Indian Institute of Technology Hyderabad, Hyderabad, Telangana 502285, India

[c)] IESL-FORTH and Materials Science and Technology Department, University of Crete, Heraklion 71110, Greece

*Corresponding Author, email: georgp@iesl.forth.gr, joshi@iitk.ac.in

**Abstract:**

Thixotropy is a phenomenon related to time dependent change in viscosity in presence or absence of flow. The yield stress, on the other hand, represents the minimum value of stress above which steady flow can be sustained. In addition, the yield stress of a material may also change as a function of time. Both these characteristic features in a material strongly influence the steady state flow curve of the same. This study aims to understand the interrelation between thixotropy, yield stress and their relation with the flow curve. In this regard, we study five thixotropic materials that show yield stress. The relaxation time of all the five systems shows power-law dependence on aging time with behaviors ranging from weaker than linear, linear to stronger than linear. Furthermore, the elastic modulus and yield stress has been observed to be constant for some systems while time dependent for the others. We also analyze the experimental behavior through a viscoelastic thixotropic structural kinetic model that predicts the observed experimental behavior of constant as well as time-dependent yield stress quite well. These findings indicate that a non-monotonic steady-state flow curve in a structural kinetic formalism necessarily leads to time-dependent yield stress, while constant yield stress is predicted by a monotonic steady-state flow curve with stress plateau in the limit of low shear rates. The present work, therefore, shows that thixotropic materials may exhibit either monotonic or non-monotonic flow curves. Consequently, thixotropic materials may show no yield stress, constant yield stress or time-dependent yield stress.

**Introduction:**

A variety of materials like concentrated suspensions and emulsions, colloidal gels, and a wide range of materials that have semisolid or paste like consistency do not reach thermodynamic equilibrium over practical timescales. As a result, such materials, in principle, undergo continuous evolution of their microstructure in order to attain progressively lower free energy as a function of time. This dynamics, also known as physical aging [1-4], induces inherent time dependency, wherein their viscosity and elastic modulus increase as a function of time. When subjected to strong deformation fields, however, the free energy of these materials increases, causing viscosity as well as elastic modulus to decrease, a process known as mechanical rejuvenation [5, 6]. This physical behavior is reminiscent of molecular glasses [7, 8], and hence this class of materials have also been termed as soft glassy materials (SGM) [4, 9]. On the other hand, thixotropy is associated with an increase in viscosity under no flow or weak flow conditions while a decrease in the same under strong flow conditions [10-15]. Consequently, combined, physical aging and mechanical rejuvenation make soft glassy materials intrinsically thixotropic [16, 17]. Owing to time dependent nature, this class of materials may show complex rheological behaviors such as time dependent yield stress [18-20], shear banding [21-25], viscosity bifurcation [26-28], overaging [29-32], delayed yielding [33-36], delayed solidification [37, 38], residual stresses [39-42], etc. Nevertheless, a material may show yield stress without being time dependent, and hence thixotropic. Furthermore, a high relaxation time viscoelastic material also takes finite time to reach steady state/equilibrium structures upon application/cessation of flow. Consequently, classic viscoelastic behavior could misleadingly be associated with thixotropy [16, 17, 43, 44]. As a result, the presence of yield stress and/or time dependency does not necessarily mean thixotropy, and hence comprehending steady state flow curve of thixotropic materials and their corresponding accurate classification is challenging. In this work we analyze different soft glassy thixotropic materials along with viscoelastic structural kinetic model towards understanding inherent flow curves and potential classification of such materials.

In soft glassy materials, constituents get arrested by entropic or enthalpic constrains such as cages or bonds formed by or with their neighbors. The resulting constrained mobility allows only limited access to the phase space, and material falls out of thermodynamic equilibrium [1, 2, 45]. The presence of yield stress is an important feature of many soft glassy materials. According to the conventional definition, when a material is subjected to a stress below the yield stress, it only undergoes elastic deformation, while for stresses above the yield stress, viscous flow is induced in the material. When the system is athermal, despite being out of thermodynamic equilibrium, it does not show any time dependency. For such materials, yield stress always remains constant [46-48]. Under the framework of generalized Newtonian fluids, the constitutive equation for a conventional yield stress fluid, can be represented as [49]:

$$\boldsymbol{\sigma} = G\boldsymbol{\gamma} \text{ for } \sqrt{(\boldsymbol{\sigma}:\boldsymbol{\sigma})/2} < \sigma_y \text{ and,}$$

$$\boldsymbol{\sigma} = \left(\frac{\sigma_y}{\dot{\gamma}} + \eta\right)\dot{\boldsymbol{\gamma}} \text{ for } \sqrt{(\boldsymbol{\sigma}:\boldsymbol{\sigma})/2} \geq \sigma_y, \quad (1)$$

where $\boldsymbol{\sigma}$ is the deviatoric stress tensor, and $\boldsymbol{\gamma}$ and $\dot{\boldsymbol{\gamma}}$ are the strain and rate of strain tensors. The second invariant of rate of strain tensor is represented as: $\dot{\gamma} = \sqrt{(\dot{\boldsymbol{\gamma}}:\dot{\boldsymbol{\gamma}})/2}$. The yield stress, elastic modulus and shear rate dependent viscosity have been represented as $\sigma_y$, $G$ and $\eta$ respectively. Typically, a yield stress fluid is modelled using Bingham plastic or Herschel-Bulkley constitutive equation. When $\eta$ is constant, the model represents the Bingham plastic constitutive equation. On the other hand, when $\eta = m\dot{\gamma}^{n_{HB}-1}$, the equation represents the Herschel-Bulkley model, where $m$ is the consistency parameter and $n_{HB}$ is the flow index [49].

Under constant pressure ($P$) and temperature ($T$), a structurally arrested out of equilibrium material with thermal constituents, spontaneously evolve to lower the Gibbs free energy state ($\bar{g}$). This process is known as physical aging. The Gibbs free energy consists of enthalpic ($\bar{h}$) and entropic contributions ($\bar{s}$): $\bar{g} = \bar{h} - T\bar{s}$. In case of physically aging materials, such as attractive emulsions, clay suspensions, colloidal gels, where constituents share predominantly attractive or repulsive interactions, microstructure formation leads to decrease in Gibbs free energy primarily due to lowering of enthalpy. On the other hand, for systems, such as the particulate colloidal glasses with hard sphere interactions, where the constituents are arrested in cages formed by neighboring particles, the decrease in Gibbs free energy is purely entropic in nature. While the systems with predominantly enthalpic interactions are reported to show significant increase in modulus during physical aging, the purely entropy driven systems show very weak or almost no increase in modulus with progress in time [6, 50-56]. Irrespective of whether physical aging has enthalpic or entropic (or both) origin, the relaxation time ($\tau$) of the material has been observed to increase as a function of time. Consequently, the viscosity (which is a product of modulus and relaxation time) of a physical aging material increases with time.

Experimentally, the dependence of the relaxation time ($\tau$) on aging time ($t_w$) can be obtained from stress relaxation or creep experiments, wherein a material is respectively subjected either to step strain or to step stress at different aging times after rejuvenation [57-59]. The stress relaxation experiments directly lead to the characteristic relaxation time, and hence obtaining its dependence on aging time is straightforward. In creep experiments, on the other hand, the dependence of the relaxation time on aging time can be obtained through creep time – aging time superposition. However, the creep time -aging time superposition works only if the physical aging does not alter the shape of the spectrum of relaxation times but affects only its mean value [34]. Consequently, creep curves at different waiting times lead to self-similar creep curves that lead to time – aging time superposition, and from the corresponding shift factor, the dependence of relaxation time on aging time can be easily obtained [60]. Furthermore, causality demands that the evolution of the relaxation time on waiting time is independent of the type of experiments employed to probe the same. Particularly, the creep compliance and the stress relaxation modulus are related to each other through a convolution relation [61], and hence an identical dependence of the relaxation time on waiting time has been reported for physically aging soft glassy materials in creep or step strain experiments [62]. For many soft glassy materials, the relaxation time has been observed to show a power law dependence on aging time [56, 58, 63-65] given by: $\tau \sim \tau_m^{1-\mu} t_w^{\mu}$, where $\tau_m$ is the microscopic timescale and $\mu$ is the power law coefficient. Depending on the value of $\mu$, the nature of aging can be categorized as hyper aging ($\mu > 1$), full aging ($\mu = 1$), or sub aging ($\mu < 1$). For some materials, on the other hand, an exponential dependence has also been reported [19, 46, 57, 66, 67] (In some of these materials, the exponential dependence of relaxation time on aging time can originate from equilibration that material undergoes immediately after a temperature or mechanical quench).

As mentioned before, physical aging involves microstructural changes leading to progressing lowering of free energy as a function of time. These might be a quite prominent microstructure build-up such as in the case of attractive colloidal gels, or very subtle structural rearrangements (often not even experimentally detectable) in repulsive colloidal glasses or jammed particle pastes. However, upon application of a strong deformation field the microstructure formed during physical aging of a material breaks or/and deforms, causing increase in free energy. Consequently, the viscosity, modulus, and relaxation time of a material decreases. This process is known as mechanical rejuvenation. Mechanical rejuvenation is partial, when application of deformation field of moderate strength does not obliterate the microstructure formed during physical aging completely, resulting only in a partial decrease in viscosity, modulus, and relaxation time [6]. Note, that this phenomenon is distinct from the overaging behavior reported in some SGMs exhibiting thixotropy, where application of a moderate magnitude deformation fields (shear rates) in turn increases the mean relaxation time instead of decreasing the same [29, 32]. Accordingly, the increase in viscosity under quiescent conditions and decrease in the same under application of deformation field together, qualify the material being called thixotropic [16, 17]. In this sense, we will retain here the classification of a material as thixotropic, based on its macroscopic phenomenological response, in letter and spirit to the conventional IUPAC definition [68], without relating it to a specific microscopic mechanism, be it of enthalpic or entropic origin.

For a microgel paste, Cloitre and coworkers [64] reported the power law dependence ($\tau \sim \tau_m^{1-\mu} t_w^{\mu}$), wherein $\mu$ is observed to be dependent on imposed creep stress. They observed $\mu$ to be unity in a limit of small stresses, which decreases with increase in stress suggesting weakening of rate of physical aging ($\mu = d\ln\tau/d\ln t_w$) or presence of partial rejuvenation. Beyond a certain stress, which they term as yield stress, physical aging completely stops, leading to $\mu = 0$. Similar behavior has also been observed for clay [57, 60, 63] and Carbopol 940 dispersions [58]. However, this yield stress is different than the classical yield stress mentioned in the Bingham or Herschel-Bulkley [Eq. (1)] model as below this yield stress the material does undergo viscous flow but with diminishing shear rate. At stresses lower than the yield stress, when $\mu > 0$, partial rejuvenation may take place along with physical aging implying a signature of thixotropy in the system. This behavior is known as viscosity bifurcation [26-28], wherein below a yield stress viscous flow does take place but with diminishing shear rate, while above the yield stress the material eventually flows with constant shear rate. More complicated behaviors such as delayed yielding [33-36], delayed solidification [37, 38], reentrant solidification [36, 69, 70] are observed wherein shear rate shows non-monotonic dependence on time under applied stress.

Recently Larson [43] proposed an important categorization of thixotropy based on separation of timescales associated with the thixotropic and viscoelastic relaxation time of a material. The characteristic thixotropic timescale is defined as that timescale which is associated with the relaxation of the viscosity under stationary conditions. According to Larson, the thixotropic response symbolizes that behavior when the characteristic thixotropic timescale is significantly larger than the viscoelastic relaxation time. A limiting case of this behavior is when relaxation time tends to zero leading to an 'ideal thixotropic' response. However, in real materials, as the microstructure builds up, the relaxation time goes on increasing. Consequently, according to Larson and Wei [44], depending on various external factors, including the strength of the applied flow field, and elapsed time, the same material may show an array of responses such as thixotropic, viscoelastic aging, viscoelastic etc. However, owing to significant viscoelasticity they keep soft glassy materials that show physical aging and rejuvenation out of the purview of thixotropy [43, 44]. While this distinction, in principle, has no logical flaw, it proposes its own taxonomy, which may not be universally acceptable. The other way to categorize thixotropic materials is to apply various adjectives to the term thixotropy by calling the same thixoviscous [71], thixoplastic [72, 73], thixotropic elasto -visco -plastic (TEVP) [74-76], etc.

Moller and coworkers [77] studied different kinds of soft materials that show yield stress and distinguished them into two categories, namely simple yield stress fluids and thixotropic yield stress fluids. A simple yield stress fluid shows yield stress but not thixotropy. On the other hand, thixotropic yield stress fluids show both yield stress and thixotropy. In order to distinguish between the two categories, it was proposed that a simple yield stress fluid that lacks thixotropy obeys the following criteria: (i) the yield stress remains constant as a function of time, (ii) the shear rate (or viscosity) obtained during increasing and decreasing shear stress must coincide leading to lack of hysteresis, (iii) the flow curve is identical irrespective of whether the measurements are carried out using imposed stress or imposed shear rate conditions and (iv) no viscosity bifurcation is observed. Based on the above criterion, it was conjectured that simple yield stress fluids always show a monotonic constitutive equation with a constant shear stress plateau in the limit of low shear rates. Conversely, they propose that thixotropic yield stress materials necessarily show a non-monotonic flow curve, wherein a decrease in stress with increasing shear rate is observed [78]. In addition, the system necessarily shows an inhomogeneous flow field leading to steady state shear banding when the imposed shear rate corresponds to a decreasing part of the shear stress.

With this background, this work attempts to classify the yield stress fluids by studying different experimental systems (clay dispersion, Carbopol dispersion and nearly hard sphere colloidal glasses) and applying a viscoelastic thixotropic model. We experimentally measure the yield stress and the relaxation time of the samples and study how both the parameters vary as a function of aging time. In addition, we deduce whether material is thixotropic or not based on a recently proposed criterion [17], which adheres to the IUPAC definition of thixotropy [68] in letter and spirit.

**Materials and Methods:**

We perform experiments on four physically aging/rejuvenating systems.: (i) a 3.5 wt. % aqueous dispersion of synthetic hectorite clay (LAPONITE RD® with the chemical formula $Na_{+0.7}[(Si_8Mg_{5.5}Li_{0.3})O_{20}(OH)_4]_{-0.7}$) at two different rest times of 14 days and 1000 days since their preparation, and (ii) hard sphere colloidal glasses at two volume fractions, $\phi$ = 0.625 and 0.595 (iii) 0.2 wt. % Carbopol dispersion. The particles of this hectorite clay have a disk-like shape with diameter of around 25 to 30 nm and thickness of 1 nm [79]. Owing to the hygroscopic nature of clay, the sample (in white powder form) is dried in an oven at 120°C for 4 h. Subsequently, 3.5 wt.% of clay is mixed with ultrapure water (resistivity 18.2 MΩcm) using IKA T50 Ultra Turrax homogenizer for 30 minutes at 10000 rpm. The dispersion is then left idle in an airtight polypropylene container. Soon after dispersing clay in water, its viscosity increases, and for concentrations greater than 1 vol % (≈ 2.5 wt. %), a soft solidlike dispersion forms which support its own weight. It has been observed that physical aging in clay dispersion is perfectly reversible upon shear over a short timescale (few days). However, it undergoes irreversible aging over long timescales (of weeks) owing to strong bond formation among the particles that cannot be broken by shear [80]. Consequently, the two clay systems with different rest times (days since preparation), 14 days and 1000 days should be considered independent systems, even though they have identical concentrations of clay.

The hard sphere colloidal suspension used in the present work consists of spherical poly(methyl methacrylate) (PMMA) particles of radius 165 nm, which are sterically stabilized with short polyhydroxystearic acid (PHSA) hairs [81, 82]. The suspension has a roughly 12% polydispersity which is enough to suppress crystallization. The particles are dispersed in an almost refractive index matched solvent squalene (viscosity 0.015 Pa.s; refractive index 1.494; and boiling temperature 421.3 °C) to minimize any remaining van der Waals attractions. Colloidal suspensions of spherical particles with polydispersity higher than 8% are known to undergo glass transition for particle volume fraction, $\phi > 0.58$. The suspension used here has a random close packing (RCP) volume fraction, $\phi \approx 0.67$, instead of $\phi \sim 0.64$ of monodispersed spheres. A single batch of particles is centrifuged to reach RCP and two successive dilutions of this batch of volume fractions $\phi = 0.625$ and $\phi = 0.595$ is prepared for rheological experiments. Note that, in comparison with the master curve created by Koumakis et al. [83] for the elastic modulus at $Pe = 0.5$ our data match very well for the two volume fractions specified above.

For the 0.2 wt.% Carbopol microgel, we borrow some data (shown in Figures 1, 2, 3, 5, and 6) from Agarwal and Joshi [58]. However, the steady state shear stress -shear rate flow curve measurements (data shown in Figure 4) and the viscosity bifurcation measurements during creep flow (data shown in Figure 7) are freshly performed for the present work. We follow the identical sample preparation and experimental protocols suggested by Agarwal and Joshi [58]. It should be noted that, the experiments of Agarwal and Joshi [58] were performed on AR-G2 rheometer while the experiments done in the present work have been performed using Anton Paar MCR 501 rheometer. We confirmed that the experimental results on these two otherwise identical samples, but analyzed using different rheometers, are similar within experimental uncertainty.

In this work, we use an Anton Paar MCR 501 stress-controlled rheometer. For clay dispersions, a serrated concentric cylinder geometry having an outer diameter of 28.605 mm with a gap of 1.12 mm is used. Shear rejuvenation of the samples is carried out for 600 s, at strain amplitude, $\gamma_{sm} = 3000\%$ for the 14 days and $\gamma_{sm} = 6000\%$ for the 1000 days rest time systems. The strain amplitude used for the aging tests and frequency sweep tests in the linear regime is $\gamma_0 = 0.1\%$. All the oscillatory experiments for clay are carried out at angular frequency $\omega = 0.63$ rad/s. For the hard sphere glasses, a serrated cone and plate geometry of diameter 25 mm and cone angle 2.7° is used. The shear rejuvenation of these samples is performed for 100 s under $\gamma_{sm} = 500\%$ and $\omega = 0.1$ rad/s, while the aging tests are carried out at $\gamma_0 = 0.1\%$ and $\omega = 1$ rad/s. All the experiments are performed at 20°C. For Carbopol dispersion, we use serrated concentric cylinder geometry, having 28.605 mm outer diameter and 1.12 mm gap thickness. The samples are mechanically rejuvenated at $\dot{\gamma} = 100$ s$^{-1}$ for 60s. The physical aging measurements are performed in oscillatory mode at $\gamma_0 = 1\%$ and $\omega = 6.3$ rad/s. All the measurements on Carbopol dispersion are carried out at 25°C. A thin layer of low-density silicone oil is added to reduce evaporation losses during measurements on the two clay dispersions and the Carbopol dispersion.

**Governing Equations:**

There are various approaches proposed in the literature to model thixotropic materials and a state of the art reviews are available on this topic [16, 43, 84]. We use a phenomenological structural kinetic model (SKM) to analyze the rheological behavior of the studied materials. This constitutes an evolution equation describing the time and deformation field dependence of a structure parameter that illustrates the instantaneous state of the structure of a material. In addition, SKM has two additional components: a constitutive equation relating stress, strain and their derivatives and expressions relating parameters of the constitutive equation, such as viscosity, modulus, etc., with the structure parameter [16, 85, 86].

In this work, we consider a structural kinetic $\lambda$-model with a viscoelastic constitutive equation. In this model, the microstructural build-up is quantified by a single structure parameter $\lambda$, such that $\lambda = 0$ represents a material devoid of any structure while in a limit of $\lambda \to \infty$, the material attains a fully developed structure. We use an evolution equation for $\lambda$ as proposed by Coussot et al. [26, 28], representing the rate of structure formation and its break-down under the application of a deformation field as a function of time ($t$) given by:

$$\frac{d\lambda}{dt} = \frac{1}{T_0} - \frac{\beta\lambda}{\eta(\lambda)}\sigma, \tag{2}$$

where $T_0$ is the characteristic time of restructuration, $\sigma$ is the second invariant of the stress tensor and $\beta$ is a scalar parameter. The first term on the right represents structural build-up or physical aging, while the second term denotes breakdown of the structure under the application of a deformation field. Some materials such as carbon black dispersion in naftenic oil, suspension of fumed silica in mixture of paraffin oil and poly(isobutylene) [86], blood [87-89] are known to undergo shear flow induced microstructural build-up. However, since shear induced aggregation is not a necessary phenomenon in thixotropic systems and none of the present experimental systems exhibits the same, we do not consider it in our model formulation. Furthermore, unlike Coussot et al. [26, 28], who use an inelastic generalized Newtonian model as the constitutive equation, we employ a time-dependent Maxwell model as the constitutive equation. The stress tensor $\boldsymbol{\sigma}$ in the time-dependent Maxwell model is related to the rate of strain tensor $\dot{\boldsymbol{\gamma}}$ as:

$$\dot{\boldsymbol{\gamma}} = \dot{\boldsymbol{\gamma}}_v + \dot{\boldsymbol{\gamma}}_e = \frac{\boldsymbol{\sigma}}{\eta(\lambda)} + \frac{d}{dt}\left(\frac{\tau(\lambda)}{\eta(\lambda)}\boldsymbol{\sigma}\right), \tag{3}$$

where $\eta$ is the viscosity and $\tau$ is the relaxation time. We relate $\eta$, $\tau$ and $\lambda$ through power law relations given by:

$$\eta = \eta_0(1 + \lambda^n), \quad \text{and} \tag{4}$$

$$\tau = \tau_0(1 + \lambda^m) \tag{5}$$

where $\eta_0$ and $\tau_0$ are respectively the viscosity and relaxation time of the thixotropic material in a completely rejuvenated or structure-less state ($\lambda = 0$), with $n$ and $m$ being the corresponding power law exponents. The expression for the viscosity is the same as that proposed by Coussot et al. [28] for their inelastic model. The expression for the relaxation time, on the other hand, is directly related to the experimental observations on various aging systems reported in the literature [46] as discussed below. Both the relaxation time of material and characteristic time of restructuration are related to the mobility of constituents of a material that are the building blocks of the structure formation. In some theoretical formulations that model structural build-up under quiescent conditions, including thixotropic models, both the timescales are shown to be related to each other [6, 8]. However, in the present case, we consider $T_0$ to be constant and $\tau$ to be a function of $\lambda$ given by Eq. (5). It is important to note that the kinetic expression of the structure parameter (Eq. (2)) considers shear rejuvenation to be proportional to stress. This suggests that, in a linear Maxwell model representation, shear rejuvenation is caused by only the shear rate associated with the dashpot (viscous component) and not the shear rate associated with the spring. Recently Joshi [85] showed that a representation of the structural kinetic model, wherein rejuvenation term involves the total shear rate (sum of shear rate associated with both spring and dashpot) leads to violation of the second law of thermodynamics. On the other hand, it was proposed that, when the rejuvenation term in a structural kinetic model depends on stress (or the viscous component of the shear rate) as described by Eq. (2), the model validates the second law of thermodynamics.

The Eqs. (2) to (5) represent the proposed linear viscoelastic constitutive model for thixotropic materials. We use the following variables to non-dimensionalize the same: $\tilde{\dot{\gamma}} = \dot{\gamma} T_0$, $\tilde{\eta} = \eta/\eta_0$, $\tilde{t} = t/T_0$, $\tilde{\sigma} = T_0 \sigma/\eta_0$, $\tilde{\tau} = \tau/\tau_0$. After non-dimentionalization, the proposed viscoelastic thixotropic model has four parameters $\beta$, $n$ and $m$. In order to fit the experimental data, which is dimensional in nature, we also need $\tau_0$, $\eta_0$ and $T_0$. However, for the flow fields considered in this work, some of the above parameters will not be needed. The discussion on how these parameters have been obtained from the experimental data is mentioned below.

**Results and Discussions:**

A. **Experimental Results**

We first probe the materials with a small amplitude oscillatory shear (SAOS) at a fixed frequency to follow the evolution of the elastic ($G'$) and viscous ($G''$) moduli in the linear regime as a function of time. Before carrying out aging tests, all the systems are shear rejuvenated, and the time of removal of shear is marked as the initiation of aging or $t_w$=0. The corresponding dynamic moduli are shown in Fig. 1. More specifically, both the clay dispersions studied herein are shear rejuvenated, as mentioned above, until their complex viscosities reach a constant steady state value. The subsequent time dependent evolutions of dynamic moduli of both the dispersions, with rest times 14 days and 1000 days, are respectively shown in Fig. 1 (a) and (b). Immediately after the cessation of shear rejuvenation, $G''$ is observed to be higher than $G'$, with the latter showing a stronger rate of increase than the former. Eventually $G'$ crosses $G''$ in such a fashion that after the crossover rate of increase in $G'$ slows down whereas $G''$ shows a maximum and subsequently decreases as a function of $t_w$. The clay gels are known to undergo irreversible type of aging [80], and hence the two systems with different rest times show some clear differences. The dispersion with a rest time of 1000 days shows higher moduli values, whereas the evolution of both moduli is faster for rest time 14 days. We also carry out dynamic frequency sweep measurements at different $t_w$ after shear rejuvenation, and the corresponding data is plotted in figures 2 (a) and (b). As the properties of clay dispersions change rapidly, the frequency is probed in the decreasing direction, and the total measurement time is kept below 200 s. It can be seen that $G'$ shows a very weak increase while $G''$ shows a weak decrease as a function of the angular frequency.

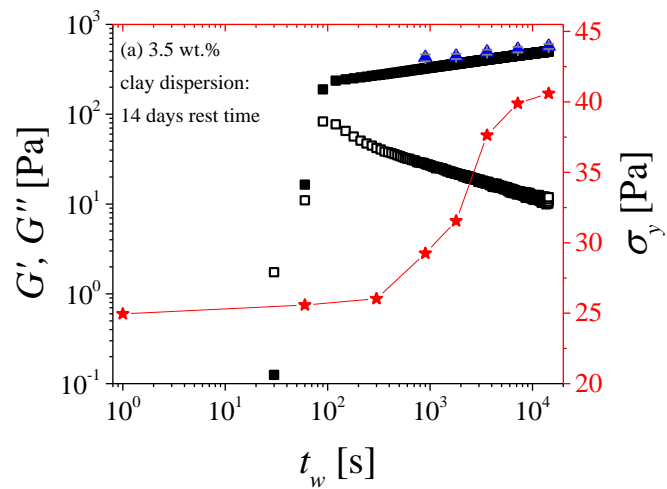

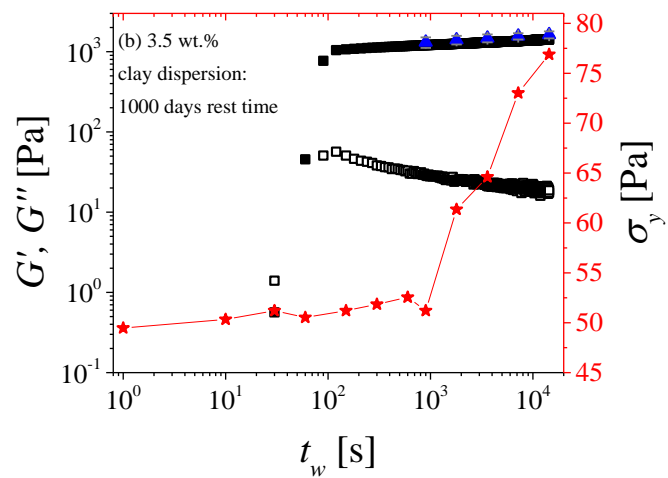

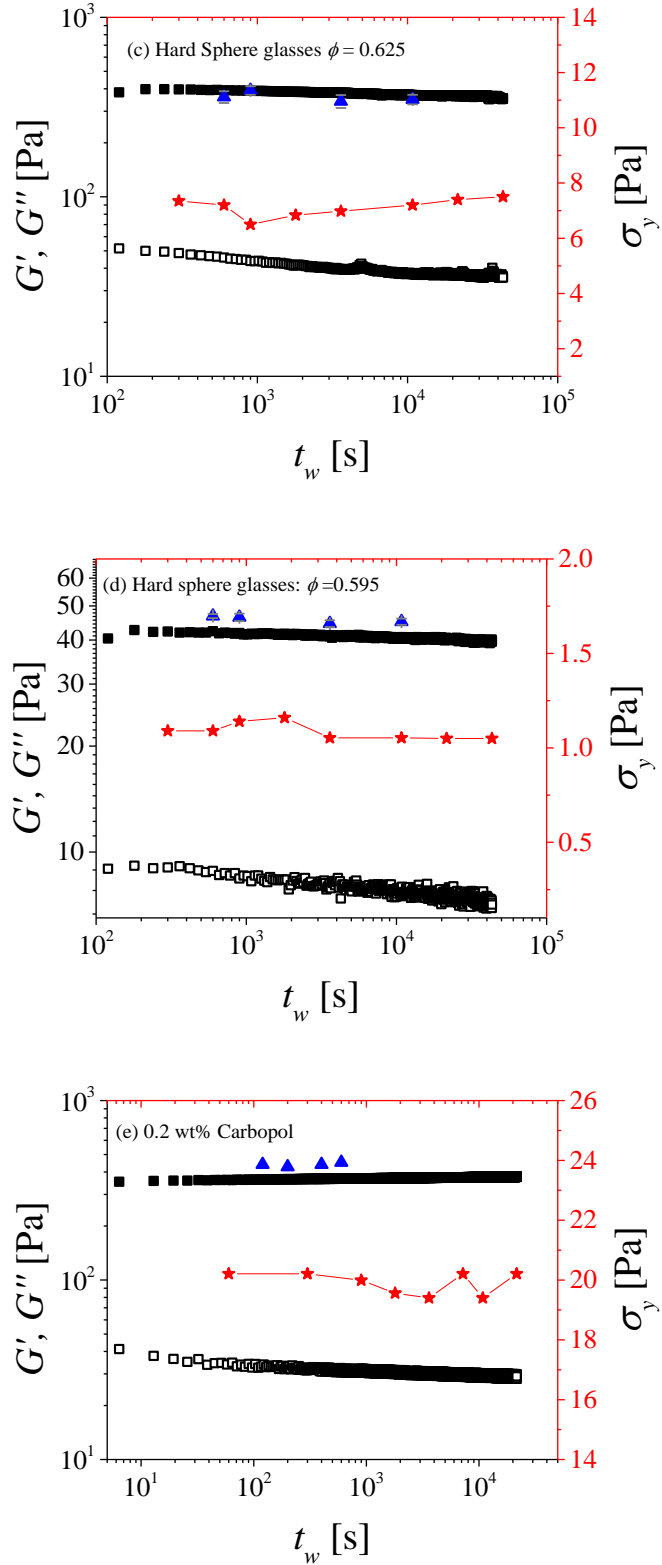

**Fig. 1** Evolution of $G'$ (squares, closed symbols), $G''$ (squares, open symbols) and $\sigma_y$ (stars) plotted as a function of aging time for 3.5 wt.% clay dispersion of rest time (a) 14 days and (b) 1000 days at $\omega=0.63$ rad/s; for hard sphere glasses with (c) $\phi=0.625$ and (d) $\phi=0.595$ at $\omega=1$ rad/s; for (e) 0.2 wt.% Carbopol at $\omega=6.28$ rad/s. The static yield stress,

$\sigma_y$ shown here is computed from the oscillatory stress sweep measurements shown in Figure 3 and Figure S1. The up triangles in each plot represent the $G'$ obtained from fitting the initial damped oscillations in strain on application of step stress, data shown in Figure S2, such that $G' \approx G$.

Similar to clay dispersions, we plot the time dependent evolution of $G'$ and $G''$ for hard sphere glasses respectively at the volume fractions of $\phi =0.625$ and $\phi =0.595$ in Figure 1 (c) and (d). In a stark difference from what is observed for clay dispersions, both glasses show a constant $G'$ and a weak decrease in $G''$ as a function of time. As expected, the dynamic moduli are observed to be higher for the volume fraction $\phi =0.625$. These observations are in line with the dynamic frequency sweep measurements presented in figures 2 (c) and (d) and earlier studies [90]. It can be seen that both the moduli increase with an increase in frequency, with $G''$ showing a stronger increase than $G'$. As expected from figure 1(c) and (d), the dynamic moduli do not show any dependence on aging time [55, 90]. In figures 1(e) and 2 (e), we replot the time evolution of dynamic moduli, and the corresponding frequency sweeps at different times for 0.2 wt.% Carbopol microgel from the work of Agarwal and Joshi [58]. We observe a very weak increase in $G'$ and a weak decrease in $G''$ as a function of $t_w$. The evolution of $G'$ and $G''$ can be seen to be frequency independent for experiments carried out at different $t_w$.

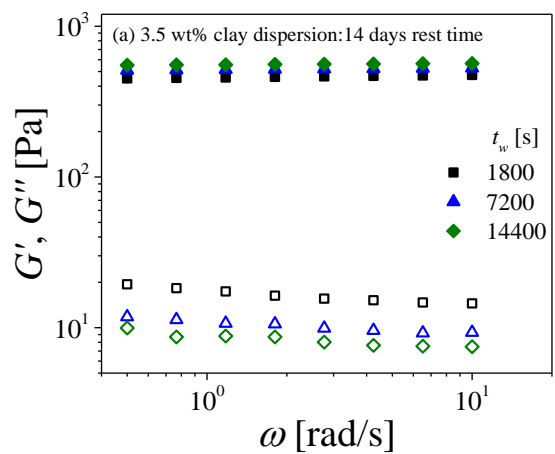

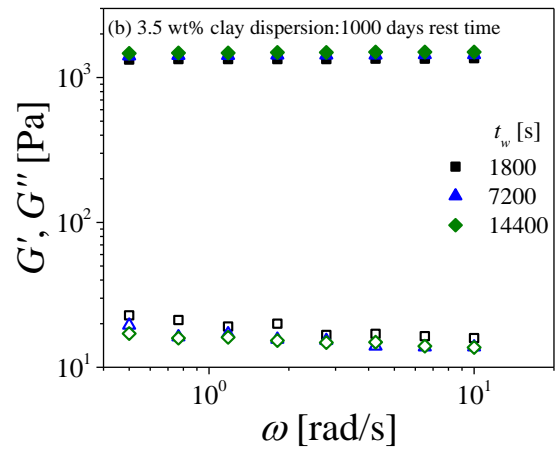

(b) 3.5 wt% clay dispersion: 1000 days rest time

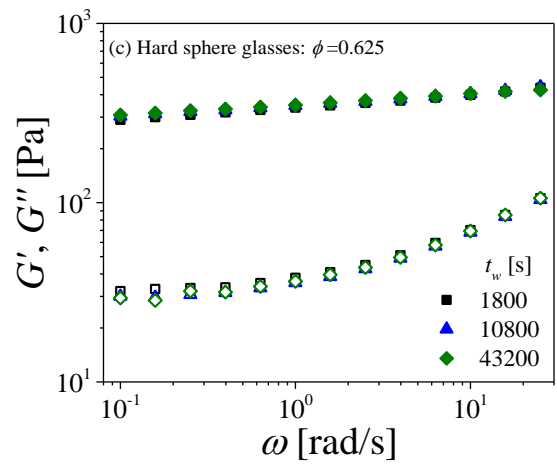

(c) Hard sphere glasses: $\phi = 0.625$

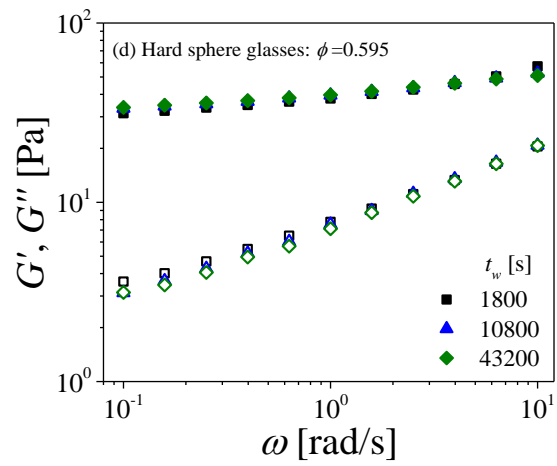

(d) Hard sphere glasses: $\phi = 0.595$

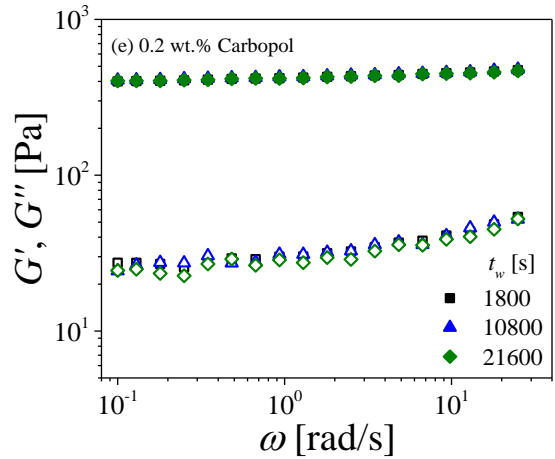

**Fig. 2** Evolution of $G'$ (closed symbols) and $G''$ (open symbols) plotted as a function of angular frequency for 3.5 wt.% clay dispersion at $\gamma$=0.1% of rest time (a) 14 days and (b) 1000 days; for hard sphere glasses at $\gamma$=0.1% with (c) $\phi$=0.625 and (d) $\phi$=0.595; for (e) 0.2 wt% Carbopol at $\gamma$=1%. The different symbols represent the different aging times since rejuvenation.

Next, we obtain the static yield stress ($\sigma_y$) of each system as a function of time elapsed since cessation of shear rejuvenation. There are many ways proposed in the literature to measure the yield stress [46, 78, 91]. In the present work, we apply an oscillatory stress ramp in the increasing direction of shear stress ($\sigma$) at a fixed frequency to obtain the yield stress. The use of increasing magnitude of oscillatory stress to estimate the yield stress, which assures a dynamic steady state at each stress magnitude, is a very convenient way to obtain the yield stress. In this procedure, when the applied magnitude of stress is much smaller than the yield stress, both $G'$ and $G''$ remains constant. As the stress approaches the yield stress, $G'$ starts to decrease while $G''$ may exhibit a maximum before decreasing or directly decrease with increasing stress amplitude. The yield stress can be obtained by employing variety of methods discussed in the literature [46, 91]. In the present work, the yield stress is obtained by considering intersection of tangents to $G'$ versus stress curve before and after the yielding. In figure 3 we plot the variation of $G'$ and $G''$ during oscillatory stress ramp measurements for the five systems at different $t_w$ since shear rejuvenation. This experiment also aids in the determination of linear regime, at which the SAOS measurements shown in Figures 1 and 2 are carried out.

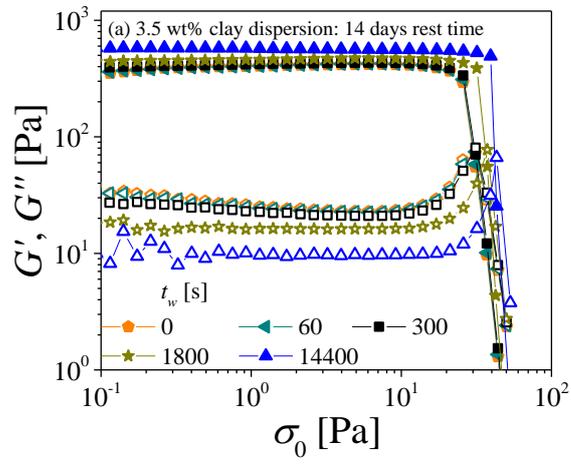

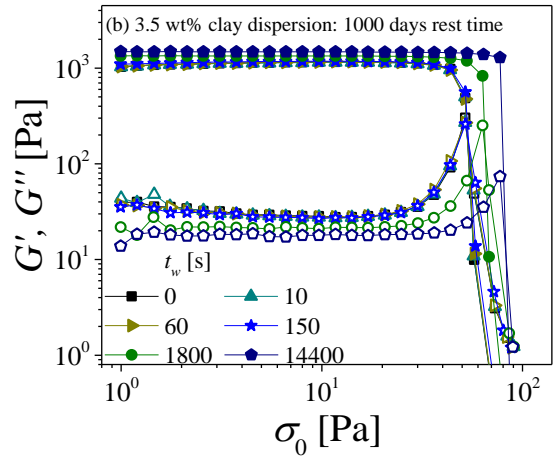

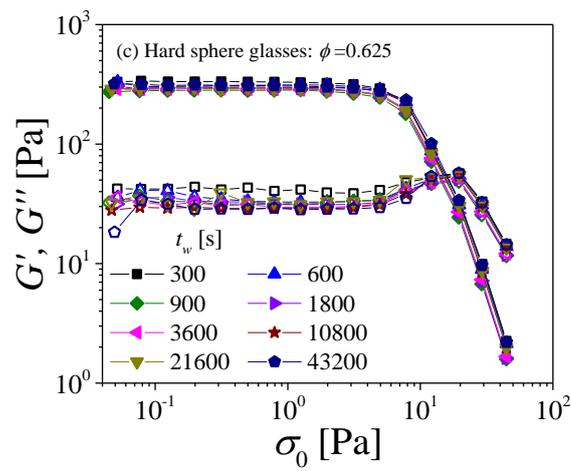

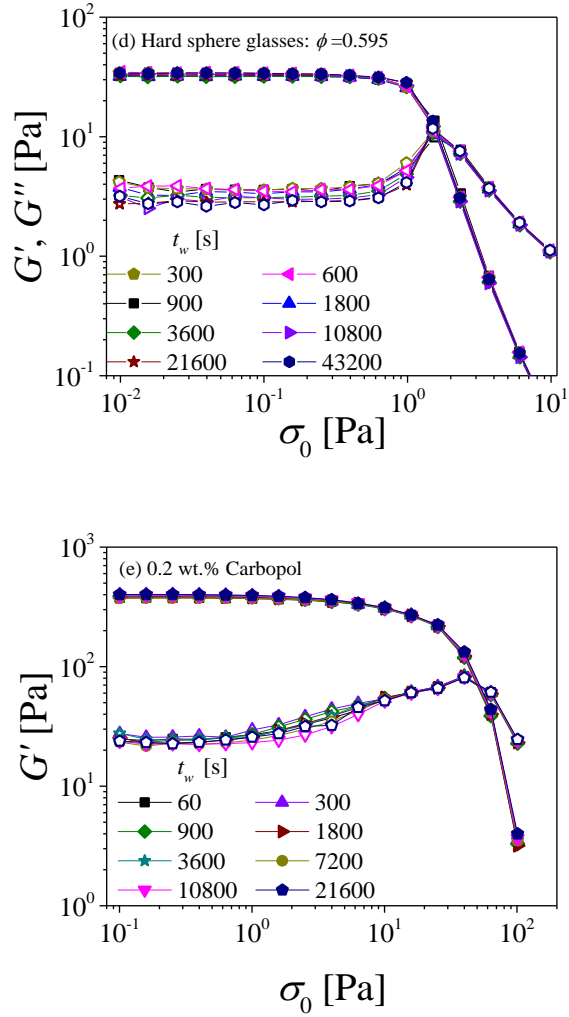

**Fig. 3** Evolution of $G'$ and $G''$ plotted as a function of increasing magnitude of oscillatory stress for 3.5 wt% clay dispersion of rest time (a) 14 days and (b) 1000 days at $\omega$=0.63 rad/s; for hard sphere glasses with (c) $\phi$=0.625 and (d) $\phi$=0.595 at $\omega$=0.1 rad/s; for (e) 0.2wt.% Carbopol at $\omega$=6.28 rad/s. The static yield stress is measured at the point of intersection of the two slopes observed in $G'$. The various symbols represent different $t_w$ as mentioned in the legends. The curves in (a) and (b) shows the time dependent static $\sigma_y$ observed in the explored range of $t_w$ for (a) 14 days ($\sigma_y \approx$ 25-40 Pa) and (b) 1000 days ($\sigma_y \approx$ 50-75 Pa) rest time of 3.5 wt.% clay dispersion. The curves in (c), (d) and (e) show the time independent static $\sigma_y$ observed in the explored range of $t_w$ for hard sphere glasses with (c) $\phi$=0.625 ($\sigma_y \approx$ 8 Pa) and (d) $\phi$=0.595 ($\sigma_y \approx$ 1 Pa) and for (e) 0.2wt% Carbopol ($\sigma_y \approx$ 20 Pa).

In figure 3 we plot the dynamic moduli as a function of the stress magnitude for clay dispersions with rest time (a) 14 days and (b) 1000 days, hard sphere glasses with volume fractions (c) $\phi=0.625$ and (d) $\phi=0.595$, and (e) 0.2 wt.% Carbopol dispersion. The corresponding static yield stresses ($\sigma_y$) as a function of aging time, $t_w$ are respectively plotted on the right-hand side ordinate of figures 1 (a) to (e). It can be seen that in the case of 14 days old clay dispersion, the $\sigma_y$ remains independent of $t_w$ at lower $t_w$. For higher $t_w$ the linear $G'$ as well as $\sigma_y$ increase with increasing $t_w$. Qualitatively similar behavior can be seen for the 1000 days old clay dispersion, with their magnitude being higher than the 14 days old clay dispersion. Interestingly for both the volume fraction of hard sphere glasses as well as for the Carbopol dispersion, the linear $G'$ and $\sigma_y$ remain nearly independent of $t_w$. As expected, the higher volume fraction of hard sphere glass exhibits a higher value of static yield stress, $\sigma_y \approx 8$ Pa compared to the lower volume fraction ($\sigma_y \approx 1$ Pa). For the 0.2 wt.% Carbopol microgel, the corresponding static yield stress, $\sigma_y$ remains constant at around 20 Pa, as shown in figure 1(e). The yield stress can also be obtained by plotting magnitude of strain as a function of the applied magnitude of stress [82, 91], as shown in the supplementary information figure S1. It can be observed that for $\sigma < \sigma_y$, the magnitude of strain shows a linear dependence on that of applied stress amplitude. However, at $\sigma \approx \sigma_y$ the slope changes, and the corresponding value of $\sigma_y$ matches very well with that obtained from procedure used in figure 3.

Very interestingly while the yield stress shows a clear increase with $t_w$ for both the clay dispersions, the strain at which yielding occurs is either independent of or very weakly dependent on $t_w$ as shown in supplementary information figure S1. The other three systems, wherein yield stress remains constant as a function of $t_w$ also shows a constant value of yield strain for the experiments carried out at different $t_w$. If we assume that the yield strain is related to the dominant length scale of the structure, this observation suggests that such length scale does not change much in all the systems explored in the present work. Such behavior is expected for the hard sphere glasses as well as the Carbopol dispersion, wherein $G'$ remains constant during physical aging. However, for the clay dispersions, $G'$ shows a significant increase. It has been established that in a gel-like system such as clay dispersions, modulus can be represented as energy density, $G' \approx c(\epsilon/b^3)$, wherein $\epsilon$ is the bond energy among the clay particles, $c$ is the proportionality constant, and $b$ is the characteristic length scale associated with the gel [60]. If the constant value of yield strain implies that the characteristic length scale of the gel remains unaffected during physical aging, an increase in $G'$ can be attributed to strengthening of the bond energy $\epsilon$.

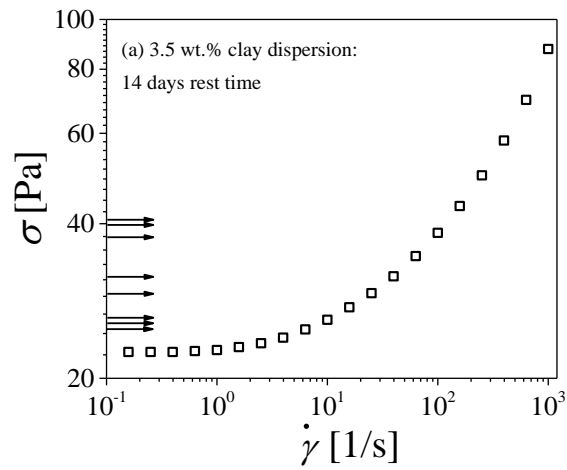

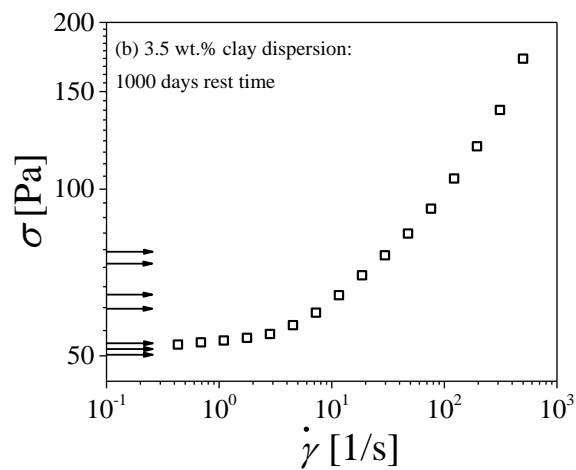

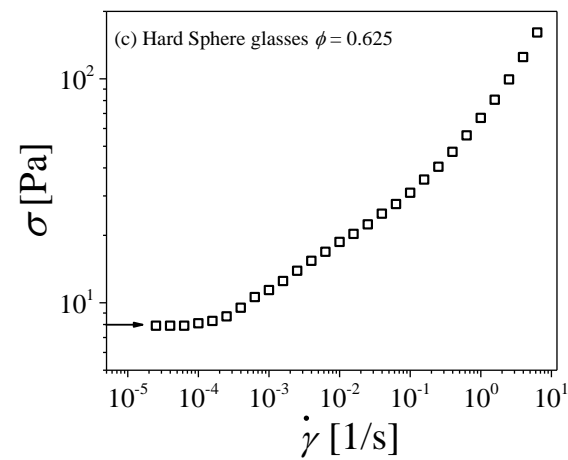

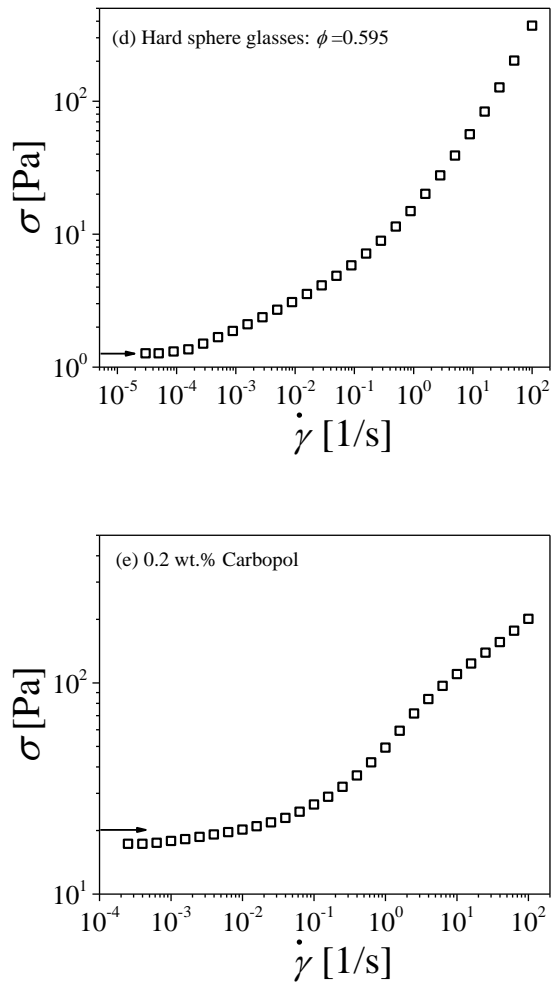

**Fig 4** The steady state flow curve for 3.5 wt.% clay dispersion (a) a 14 days rest system and (b) a 1000 days rest time system; for hard sphere glasses with volume fractions, (c) $\phi$=0.625 and (d) $\phi$=0.595; for (e) 0.2 wt.% Carbopol. The arrows shown on the ordinate are associated with the yield stress values mentioned in Fig. 1. The yield stress associated with clay dispersions shown in (a) and (b) increases with time and hence has been depicted by multiple arrows, while for the other systems, the yield stress remains constant and has been described by a single arrow. As discussed in the text, for systems that show time dependent yield stress there is more to the steady state flow curve (shown in (a) and (b)) than what can be obtained using conventional rheometer.

We also obtain the steady state flow curves associated with the five experimental systems by subjecting the same to a decreasing shear rate ramp after the rejuvenation. We wait at each point long enough to have material undergo several strain units of deformation ($t\dot{\gamma} > 1$) and ensure the steady state. In Figure 4 we plot the shear stress as a function of applied shear rate. The flow curve of all the five systems exhibits a plateau in the shear stress in a limit of small shear rates, which is suggestive of a presence of yield stress. In the same plot, we represent the yield stress obtained from the oscillatory shear stress sweep experiments by arrows on the ordinate axis. It can be seen that, for the three systems, namely two hard sphere glasses and Carbopol dispersion, the constant yield (time independent) stress obtained from oscillatory experiments matches well (within experimental uncertainty) with the plateau value of shear stress shown in Figure 4. The two clay dispersions, on the other hand, show time dependent yield stress, wherein the minimum value of the same matches well (within experimental uncertainty) with the plateau value of shear stress shown in Figure 4. This suggests that there is more to flow curve when yield stress depends on time than what is depicted by the steady state flow curve obtained from a rheometer. It is known that the region of the flow curve where stress decreases with shear rate is linearly unstable and such behavior cannot be captured by conventional rheometer procedure [92]. In a rate controlled flow field, negative slope associated with shear stress – shear rate flow curve leads to inhomogeneous flow field resulting in steady state shear banding [23, 93]. In this case, the flowing band usually flows with shear rate associated with minimum value of shear stress. Consequently, the conventional rheometer only shows a shear stress plateau associated with minimum value of shear stress in the flow curve [6, 23]. We elaborate on this aspect more while analyzing origin of time dependent yield stress below.

We now obtain and analyze the dependence of relaxation time on aging time for all the five systems explored in this work. As mentioned in the literature, such dependence is obtained by subjecting a material to creep tests at different aging times ($t_w$). We plot the creep curve for 3.5 wt.% clay dispersion of 14 days rest time in supplementary information Figure S2 for an applied stress of 5 Pa. The strain induced in a material as a function of creep time (time elapsed since the application of step stress) initially shows creep ringing for times of $\mathcal{O}(1)$ s. This is observed for all the five systems explored and originates from a coupling of the instrument inertia with the viscoelasticity of studied materials [54, 94, 95]. Such creep ringing has been analyzed in literature by applying Newton's second law of motion to the rheometer system by assuming the viscoelastic material to follow the Maxwell-Jeffreys constitutive model [96]. The corresponding time evolution of strain upon application of step stress of magnitude $\sigma_0$ is then given by [95]:

$$\gamma(t) = \sigma_0 \left\{ \frac{t}{\eta_2} - B + e^{-At} \left[ B \cos \bar{\omega} t + \frac{A}{\bar{\omega}} \left( B - \frac{1}{A\eta_2} \right) \sin \bar{\omega} t \right] \right\} \qquad (6)$$

where, $A = (aG + \eta_1 \eta_2)/(2a(\eta_1 + \eta_2))$; $B = (a(\eta_1 + \eta_2)/\eta_2 G)((2A/\eta_2) - (1/a))$; and $\bar{\omega} = \sqrt{(\eta_2 G/a(\eta_1 + \eta_2)) - A^2}$. Here $a$ is the moment of inertia of the moving part of the rheometer, while the parameters $\eta_1$, $\eta_2$, and $G$ are associated with the Maxwell-Jeffreys constitutive model. In the Maxwell-Jeffreys model, the Kelvin Voigt element ($\eta_1$ and $G$) is in series with the dashpot ($\eta_2$) [95]. In some formulations, viscosity of the dashpot is assumed to be so high that (limit of $\eta_2 \to \infty$), Maxwell-Jeffreys constitutive model reduces to Kelvin Voigt model that also predicts inertial oscillations [54, 94]. We fit Eq. (6) to the experimental data on each of the five experimental systems studied in this work at different $t_w$ and all stresses smaller than yield stress ($\sigma_0 < \sigma_y$). The average value of the estimated $G$ has been plotted as a function of $t_w$ in Figure 1. The values of $G$ obtained from this procedure as a function of $t_w$ are nearly equal to $G'$ as shown in Figure 1. It should be noted that, unlike the structural kinetic model wherein we use time dependent single mode linear Maxwell model, in Eq. (6) we use Maxwell-Jeffreys constitutive model with constant parameters. In this case, a spring is in parallel with a dashpot, which in addition to inertia, is a necessary ingredient to model creep ringing. Since creep ringing occurs immediately after a system is subjected to a step stress and over a brief time span, we assume its properties remain constant.

In the inset of Fig. 5(a) we plot the compliance ($J(t)$) induced in 14 days old clay dispersion for applied stress of $\sigma = 5$ Pa scaled with the corresponding modulus ($G(t_w)$) obtained from Eq. (6) at different $t_w$. It can be seen that at any creep time ($t - t_w$) the normalized compliance decreases with increase in $t_w$. Furthermore, the creep curves become less steep with increasing $t_w$. The very fact that, the creep curves do not overlap on to each other suggests that the material does not follow time translational invariance (TTI), and additional dependence on $t_w$ is observed ($J = J(t - t_w, t_w)$). This behavior has been reported by variety of soft glassy materials that undergo physical aging, and its origin is the evolution of relaxation time ($\tau$) and modulus (if any) as a function of $t_w$. Owing to additional $t_w$ dependence leading to breakdown of TTI, such materials do not follow the Boltzmann superposition principle. Customarily, for such materials, the material clock is transformed from the real time domain ($t$) to the effective time domain $\xi(t)$, wherein the real time is normalized by an ageing time, dependent relaxation time $(\tau(t_w))$. Therefore, the relaxation time remains constant in the effective time domain. The effective time is defined as [7, 97]:

$$\xi(t) = \tau_1 \int_0^t \frac{dt_w}{\tau(t_w)} \tag{7}$$

where, $\tau_1$ is the constant relaxation time associated with the effective time domain. Consequently, the Boltzmann superposition principle in the effective time domain can be expressed as [7, 97]:

$$\gamma(\xi) = \int_{-\infty}^{\xi} J(\xi - \xi_w) \frac{d\sigma}{d\xi_w} d\xi_w \tag{8}$$

where, $\xi_w = \xi(t_w)$ is the effective time at which the deformation field is applied and $J(\xi - \xi_w)$ is the creep compliance in the effective time domain. Since in the effective time domain, the compliance is a function of only the effective time elapsed since the application of the deformation field, a material follows an effective TTI in the effective time domain. However, in order to express the effective time, one must have prior knowledge of the functional dependence of the relaxation time on aging time $(\tau = \tau(t_w))$. There are various ways to obtain this dependence as mentioned in the literature. However, most of the soft glassy materials studied in the literature follow a power law dependence of their relaxation time on aging time $(t_w)$ given by [57, 61, 62]:

$$\tau = A_1 \tau_m^{1-\mu} t_w^{\mu} \tag{9}$$

where, $A_1$ is a constant, $\tau_m$ is the microscopic relaxation time of a material and $\mu$ is the power law coefficient. The incorporation of Eq. (9) in Eq. (7) and considering $\tau_1 = \tau_m$ leads to:

$$\xi - \xi_w = \tau_1 \int_{t_w}^{t} \frac{dt'}{\tau(t')} = \frac{\tau_1^{\mu}}{A_1} \left[ \frac{t^{1-\mu} - t_w^{1-\mu}}{1-\mu} \right] \tag{10}$$

In Fig. 5(a) we plot the $J(t)G(t_w)$ as a function of $\frac{t^{1-\mu} - t_w^{1-\mu}}{1-\mu}$ at different $t_w$ for the 14 days old clay dispersion. It can be seen that all the creep curves superpose, leading to a master curve for $\mu = 2.35 \pm 0.04$. The existence of such a superposition supports the usage of the power law dependence of relaxation time on aging time given by Eq. (9). In this superposition $G(t_w)$ acts as a natural vertical shift factor as discussed elsewhere [61]. In the inset of Figs. 5 (b), (c), (d) and (e) we plot $J(t)G(t_w)$ respectively for the 1000 days old clay dispersion ($\sigma = 20$ Pa), the hard sphere glass with $\phi = 0.625$ ($\sigma = 0.5$ Pa), the hard sphere glass with $\phi = 0.595$ ($\sigma = 0.5$ Pa), and Carbopol dispersion ($\sigma = 3$ Pa) as a function of creep time $(t - t_w)$. The respective values of stresses are chosen such that $\sigma/\sigma_y \leq 0.5$ for all the five systems. The normalized compliance curves obtained at different $t_w$ superpose when plotted in the effective time domain. The corresponding superpositions have been shown in Figs. 5 (b) (c), (d) and (e) for the 1000 days old clay dispersion ($\mu = 1.13 \pm 0.05$), the hard sphere glass suspension with volume fraction 0.625 ($\mu = 0.9 \pm 0.03$), the hard sphere glass suspension with volume fraction 0.595 ($\mu = 0.55 \pm 0.01$), and Carbopol dipsersion ($\mu = 0.97 \pm 0.01$) respectively.

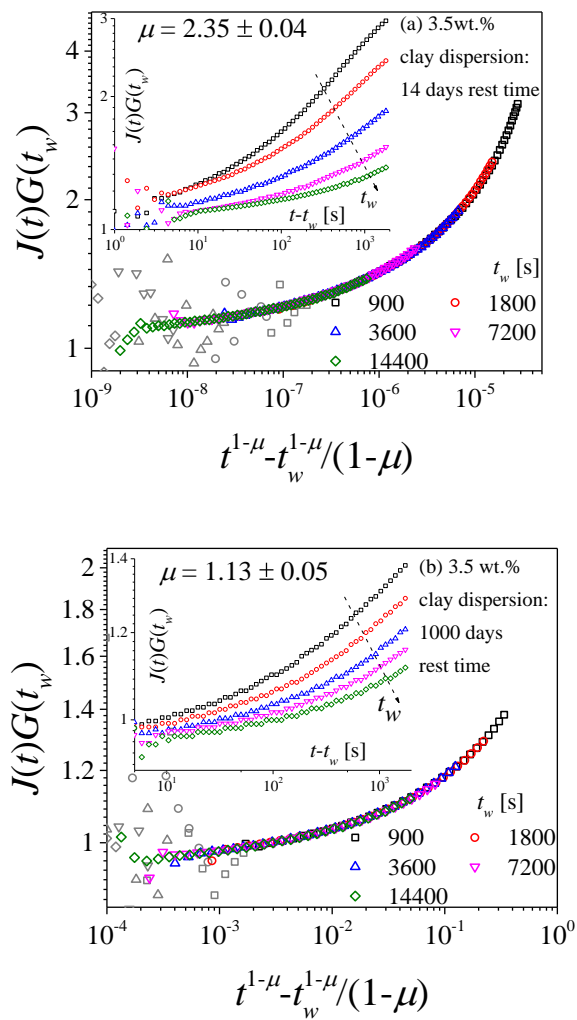

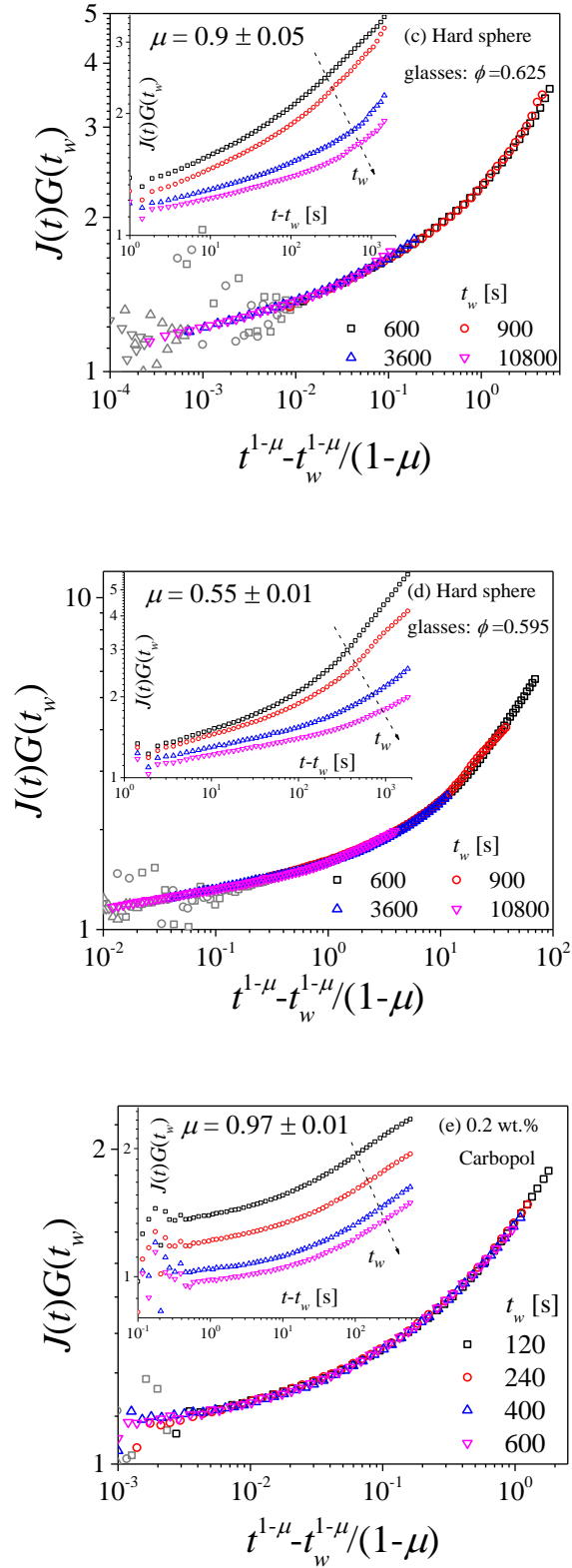

**Fig 5** Time aging time superposition of creep compliance obtained as a function of $t_w$ for 3.5 wt% clay dispersion (a) a 14 days old system at $\sigma$= 5 Pa ($\mu$ = 2.35) and (b) a 1000 days old system at $\sigma$= 20 Pa ($\mu$ = 1.13); for hard sphere glasses with (c) $\phi$=0.625 at $\sigma$= 0.5 Pa ($\mu$ = 0.9) and (d) $\phi$=0.595 at $\sigma$= 0.5 Pa ($\mu$ = 0.55); for (e) 0.2wt% Carbopol at $\sigma$= 3 Pa ($\mu$ =

0.97). The various symbols represent the different $t_w$ as mentioned in the legends. The method to obtain the vertical shift factor $G(t_w)$ is shown in supplementary information figure S2. In the insets, the corresponding data for normalized compliance is shown as a function of creep time $(t - t_w)$. The arrows guide the eyes in the increasing direction of $t_w$.

We carry out the procedure of obtaining creep curves at different waiting times, whose results are shown in Fig. 5 at different values of stress in order to obtain the dependence of $\mu$ on applied stress during creep. We observe that for two systems, namely the hard sphere glasses with volume fraction of 0.595 and the 0.2 wt.% Carbopol dispersion negative strain values are detected for low values of stress. Such negative strain for applied positive stress is known to be because of continuing recovery of strain after shear rejuvenation. In the limit of linear viscoelasticity, the response (strain) associated with the independent application of impetus (stress) is additive. Consequently, in the present case, the strain exclusively associated with the applied creep stress $\gamma(t - t_w, t_w)$ can be obtained by accounting for the recovered strain for zero stress at different waiting times $\gamma_0(t - t_w, t_w)$ with the knowledge of strain as recorded by the rheometer $\gamma_1(t - t_w, t_w)$, using: $\gamma_1(t - t_w, t_w) = \gamma(t - t_w, t_w) + \gamma_0(t - t_w, t_w)$. The detailed discussion about the origin of this phenomenon and the procedure to obtain the true strain associated with the applied creep stress has been discussed by Agarwal and Joshi [58], from which the rheological data of 0.2 wt. % Carbopol dispersion has been reproduced.

For the two clay dispersion systems with rest times of 14 and 1000 days, it has already been reported that yield stress depends on waiting time $(t_w)$. Consequently, for an applied stress $\sigma$, those creep curves obtained at such $t_w$ for which $\sigma$ is close to $\sigma_y(t_w)$, will not participate in the superposition. In the supplementary information in Figures S3 and S4, we plot creep curves obtained at various $t_w$ (as the inset) and their superposition in the effective time domain respectively at rest time of 14 and 1000 days. It can be seen that for day 14 dispersion, whose minimum yield stress is 25 Pa according to Fig 1(a), the creep curves obtained up to the stress of 10 Pa participate in the superposition, irrespective of the value of waiting times. However, for the applied stress of 20 Pa, the creep curves associated with $t_w \leq 1800\ s$ do not participate in superposition as the value of stress is closer to the yield stress that alters the shape of the relaxation time spectrum. For the $\sigma > 30$ Pa and beyond, the system shows delayed yielding for higher waiting times. However, after a long creep time, these curves are expected to collapse, leading to $\mu = 0$. Qualitatively similar behavior has also been observed for the clay dispersion with rest time of 1000 days plotted in supplementary information Figures S4. For the remaining three systems, since the yield stress does not depend on waiting time, all creep curves participate in the superposition over the explored values of stresses, as also shown by Agarwal and Joshi [58].

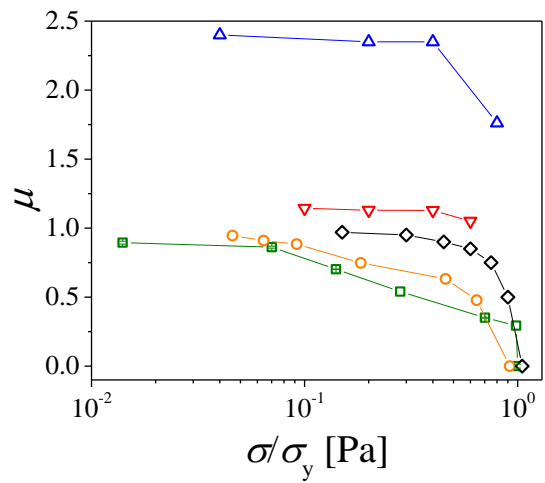

**Fig 6** Power law exponent $\mu$, used at different stresses to obtain the time aging time superposition, as a function of stress normalized by yield stress ($\sigma_y$) of the materials for the systems. The symbols represent different systems: for 3.5 wt.% clay dispersion: up triangles for rest time of 14 days and down triangles for rest time of 1000 days old; for hard sphere glasses: squares for $\phi$=0.625 and circles for $\phi$=0.595; and rhombuses for 0.2 wt.% Carbopol system.

In Fig. 6, we plot $\mu$ as a function of the stress normalized by the yield stress of the materials. For the case of the two clay dispersions, we use the minimum value of yield stress for normalization. In general, it can be seen that the power law coefficient $\mu$ remains constant in a limit of small stresses and decreases with increasing stress. Eventually, $\mu$ drops and approaches zero as the stress approaches the yield stress, and at that stress, the strain induced in material does not depend on $t_w$. The value of $\mu$ in the limit of very small stresses ($\sigma \ll \sigma_y$) has a special significance as, in this limit, the applied deformation field does not influence the evolution of relaxation time, and the value of $\mu$ can be taken as that associated with the physical aging of the material under quiescent conditions. In this sense, Fig. 6 suggests that both the clay systems show a hyper aging ($\mu > 1$) kind of behavior. The system with a rest time of 14 days shows a very strong power law dependence where $\mu = 2.35$, whereas the system with rest time of 1000 days shows a much weaker dependence close to linear as $\mu = 1.13$. The decrease in $\mu$ as a function of rest time for the studied clay dispersion is due to the slowly varying microstructural dynamics of clay particles in aqueous media over a prolonged period of time. The details of this behavior have been discussed elsewhere [60]. On the other hand, the two hard sphere glasses with different volume fractions and the Carbopol microgel show a simple to sub aging behavior ($\mu \leq 1$) where $\mu$ varies in the range of 0.9 to 1. The Carbopol dispersion, whose results have been borrowed from Agarwal and Joshi [58], is observed to show a simple aging behavior with $\mu \approx 1$.

Interestingly, the procedure employed to obtain the dependence of the relaxation time on aging time has also been prescribed by Agarwal et al. [17] to assess the thixotropic character of a material. The authors suggested that when creep experiments are performed at different $t_w$ after the stoppage of preshear (shear rejuvenation) if the compliance ($J(t - t_w, t_w)$) at a fixed creep time ($t - t_w =$ constant) decreases with $t_w$, a material can be termed as thixotropic. Accordingly, in the insets of Fig. 5, if we consider the compliance at a constant $t - t_w$ (vertical line) and plot it with $t_w$, we shall observe that it decreases with $t_w$. Therefore, as per the criterion proposed by Agarwal et al. [17], which adheres to the IUPAC definition of thixotropy [68], all the systems studied in this work are thixotropic in nature.

The behavior shown in Figs. 5 and 6 also necessarily implies the presence of a viscosity bifurcation. Furthermore, Joshi and Petekidis [46] reported that for a stress dependent power law coefficient $\mu = \mu(\sigma)$, the strain rate associated with system evolves according to $\tilde{\dot{\gamma}} \sim \tilde{t}^{-\mu(\sigma)}$. This suggests that $\dot{\gamma}$ must gradually decrease with time under the application of stress for $\mu > 0$. Since, for $\sigma \geq \sigma_y, \mu = 0$, a continuous flow with finite shear rate gets induced in a material. This analysis, therefore, clearly suggests that viscosity bifurcation must be present for all the investigated systems, with the point of bifurcation being the yield stress, irrespective of whether it is constant or time dependent.

Nonetheless, we experimentally investigate the various materials for their viscosity bifurcation behavior.

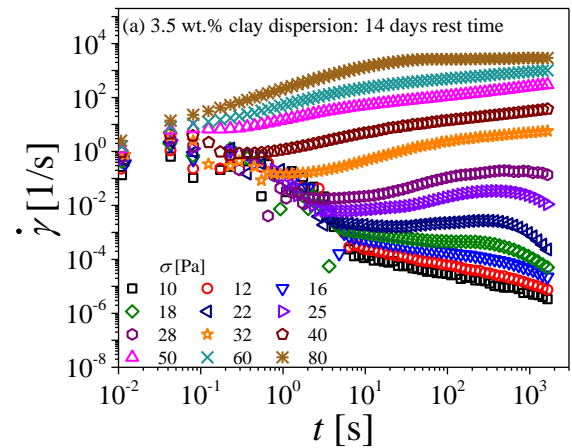

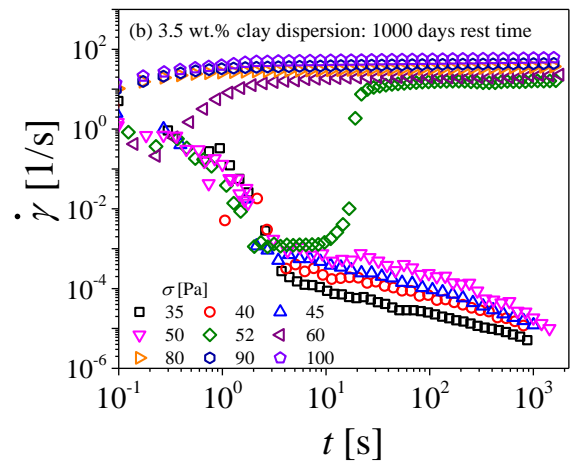

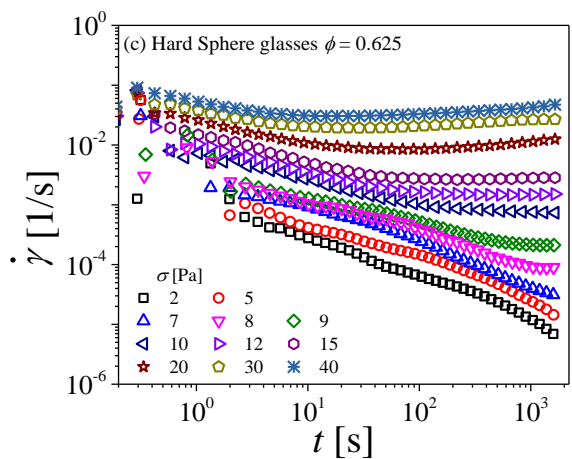

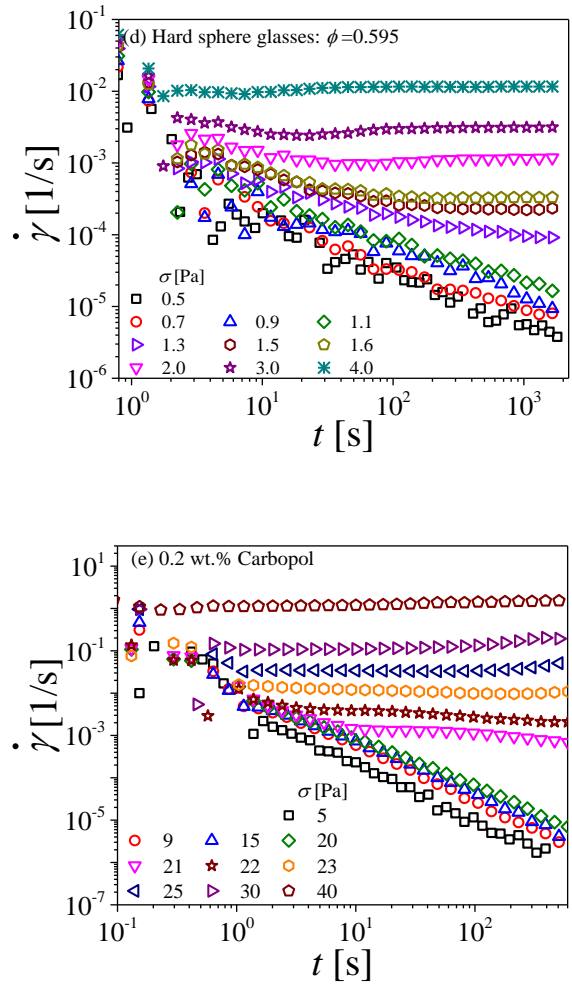

**Fig 7** Evolution of shear rate is plotted as a function of time upon application of constant stress for 3.5 wt.% clay dispersion (a) a 14 days old system at $t_w = 900$ s ($\sigma_y \approx 27$ Pa) and (b) a 1000 days old system at $t_w = 900$ s ($\sigma_y \approx 50$ Pa); for hard sphere glasses with (c) $\phi$=0.625 at $t_w = 900$ s ($\sigma_y \approx 8$ Pa) and (d) $\phi$=0.595 at $t_w = 900$ s ($\sigma_y \approx 1$ Pa), and for (e) 0.2 wt.% Carbopol at $t_w = 120$ s ($\sigma_y \approx 20$ Pa). The various symbols represent the different $\sigma$ as mentioned in the legends.

In Figure 7 we plot the temporal evolution of shear rate during creep measurements performed at different stresses for the five systems. As mentioned previously, all the systems were shear rejuvenated and aged for a fixed time prior to any measurement. For the clay dispersion with 14 days rest time and an aging time of $t_w$=900 s, Figure 7(a) shows that at applied stresses in the range of $\sigma = 10 - 16$ Pa, which is much below the static yield stress of the system ($\sigma_y \approx 27$ Pa), shear rate dramatically decreases by six orders of magnitude within the experimental observation time. On increasing the applied magnitude of stress, in the range of $18 \leq \sigma < 28$ Pa, the evolution of shear rate shows a peak before eventually decreasing with time. Finally, at higher magnitudes of applied stresses, $\sigma \geq 28$ Pa, the shear rate is always observed to increase monotonically with time and eventually attains a plateau or a steady state. Accordingly, the evolution of shear rate can be segregated in three regimes: (I) for smaller stresses, below 18 Pa, the evolution is solely driven by physical aging or structural build up; (II) for intermediate stresses in the range of $18 \leq \sigma < 28$ Pa, the behavior is a consequence of competition between physically aging relaxation modes and rejuvenating relaxation modes [98]; (III) for higher stresses, $\sigma \geq 28$, the evolution is solely driven by rejuvenation or structural break-down. This behavior, where a system flows with a continuously decreasing shear rate (viscosity increases) at lower magnitudes of applied stresses, but at larger stresses eventually shows increase in shear rate leading to a steady state (viscosity decreases), is known as viscosity bifurcation in the literature. In Figure 7 (b) we plot the evolution of shear rate during creep tests carried out on clay system with 1000 days rest time at an aging time of $t_w$=900 s. However, in this case only two shear rate regimes are observed while demonstrating viscosity bifurcation. The two regimes are widely separated by a steep increase in shear rate owing to delayed yielding near the static yield stress of 50 Pa. The Figures 7 (c) and (d) show the results associated with the creep tests in the vicinity of static yield stress, on the two hard sphere glasses of volume fractions $\phi$=0.625 and $\phi$=0.595 respectively. Both the systems exhibit a flow with diminishing shear rate at small stresses. Interestingly, with a minute increase in applied stress, the shear rate ultimately attains a plateau. The time at which shear rate attains a steady state, decreases with increasing magnitude of applied stress, suggesting viscosity bifurcation. Qualitatively similar viscosity bifurcation behavior has also been observed for the Carbopol microgel as shown in Figure 7(e), where shear rate progressively decreases with time for stresses applied below the yield stress but reaches a steady state when applied stress is higher than the yield stress.

In order to get further insight into the observed behavior, particularly the dependence of yield stress, modulus, and relaxation time on the waiting time, we analyze the experimental data from the point of view of the structural kinetic model. with viscoelastic constitutive equation described before.

## B. Model Predictions:

We first analyse the predictions of the structure kinetic model in the limit of steady state. Here we denote the steady state flow conditions with subscript *ss*. Since, according to the Maxwell model, the elastic component of strain remains constant in the limit of the steady state, we get $\tilde{\dot{\gamma}}_{ess} = 0$. Consequently, in the limit of the steady state the total shear rate becomes equal to the viscous shear rate, i.e., $\tilde{\dot{\gamma}}_{ss} = \tilde{\dot{\gamma}}_{vss}$. Accordingly, the steady state form of Eq. (2) becomes, $\lambda_{ss} = \frac{1}{\beta \tilde{\dot{\gamma}}_{vss}} = \frac{1}{\beta \tilde{\dot{\gamma}}_{ss}}$. On the other hand, noting the constitutive equation given by Eq. (3) and the relationship between structure parameter $\lambda$ and material properties $\eta$ and $\tau$ given by Eqs. (4) and (5), we get:

$$\tilde{\sigma}_{ss} = \left(1 + \left(\beta \tilde{\dot{\gamma}}_{ss}\right)^{-n}\right) \tilde{\dot{\gamma}}_{ss}. \tag{11}$$

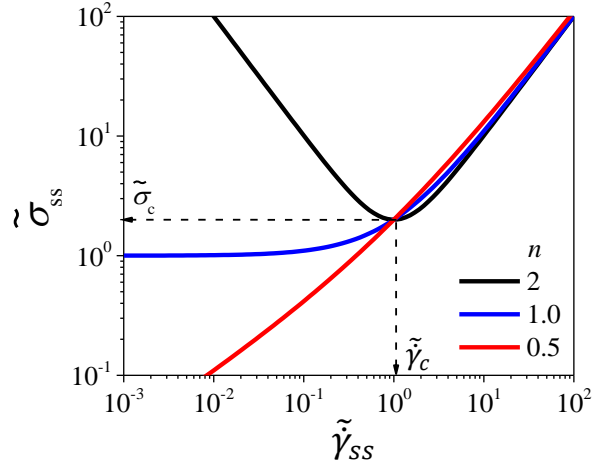

**Fig. 8** Steady state flow curve for the structure kinetic model with viscoelastic constitutive equation represented by Eq. (10) at different values of the power law coefficient $n$ relating viscosity $\eta$ and structure parameter $\lambda$ through Eq. (4) at $\beta = 1$.

In Fig. 8 we plot $\tilde{\sigma}_{ss}$ as a function of $\tilde{\dot{\gamma}}_{ss}$ for different values of $n$ at $\beta = 1$. For $n \leq 1$, the steady state relationship between $\tilde{\sigma}_{ss}$ and $\tilde{\dot{\gamma}}_{ss}$ is monotonic, while for $n > 1$, $\tilde{\sigma}_{ss}$ decreases with $\tilde{\dot{\gamma}}_{ss}$ below a critical shear rate $\tilde{\dot{\gamma}}_c$ (with corresponding stress to be $\tilde{\sigma}_c$ and structure parameter $\lambda_c = 1/\tilde{\dot{\gamma}}_c$). A special case is $n = 1$, for which $\tilde{\sigma}_{ss}$ shows a plateau in a limit of small $\tilde{\dot{\gamma}}_{ss}$, that indicates a constant (time-independent) yield stress, although the behavior is indeed thixotropic. For values of $n > 1$, the model naturally shows a time dependent yield stress as discussed below. Consider a case in which, at time $t = 0$, the structure parameter has a value $\lambda = \lambda_0$, that is indicative of the extent of shear rejuvenation in a real experimental system: the smaller the value of $\lambda_0$, the more intense the effect of shear on the structural breakdown. If a model is subjected to $\tilde{\dot{\gamma}} = 0$ at $t = 0$, $\lambda$ will evolve linearly given by $\lambda - \lambda_0 = \tilde{t}$, as per to Eq. (2) and the system will take time equal to $\tilde{t}_c = \lambda_c - \lambda_0$ to reach $\lambda = \lambda_c$. At any point in time, if the applied stress is below $\tilde{\sigma}_c$, the material will flow with diminishing shear rate leading eventually to a complete cessation of flow. Consequently, the minima of the shear stress-strain rate flow curve that is $\tilde{\sigma}_c$, can be considered as the minimum value of the yield stress associated with the system. If the evolved $\lambda$ attains a value of $\lambda_1$ such that $\lambda_1 > \lambda_c$, for any applied stress $\tilde{\sigma}$, the instantaneous $d\lambda/d\tilde{t}$ will be positive for $[(1 + \lambda_1^n) - \beta\lambda_1\tilde{\sigma}] > 0$ as per to Eq. (2). Consequently, the viscosity will increase continuously causing the shear rate to decrease, leading to eventual flow stoppage. On the other hand, for $[(1 + \lambda_1^n) - \beta\lambda_1\tilde{\sigma}] < 0$, $\lambda$ as well as the viscosity will change in such a fashion that the shear rate will ultimately reach a finite steady state value. As a result, the non-monotonic nature of the flow curve will lead to a time dependent yield stress given by:

$$\tilde{\sigma}_y = \tilde{\sigma}_c \qquad \text{for } \lambda \leq \lambda_c \text{ (or } \tilde{t} \leq \tilde{t}_c\text{)} \qquad \text{and}$$
$$\tilde{\sigma}_y = \frac{1}{\beta}\frac{(1 + (\tilde{t} + \lambda_0)^n)}{(\tilde{t} + \lambda_0)} \quad \text{for } \lambda > \lambda_c \text{ (or } \tilde{t} > \tilde{t}_c\text{)} \tag{12}$$

Therefore, the yield stress will remain constant for $\tilde{t} \leq \tilde{t}_c$ at the value associated with the minima of the flow curve. For $\tilde{t} > \tilde{t}_c$ the yield stress will increase monotonically as shown by Eq. (12). Similar behavior has also been reported for a fluidity model proposed by Joshi [6].

Now we discuss the estimation of the model parameters required to realistically represent the experimental response of the systems explored in this work. The model has three fitting parameters, $n$, $m$ and $\beta$. Firstly, in order to obtain $n$, we need to fit the experimental viscosity data to Eq. (4). The viscosity of the various materials studied here is given by $\eta = G\tau$. The model considers the relaxation time given by $\tau = \tau_0(1 + \lambda^m)$. However, under quiescent conditions, we have $\lambda = \tilde{t}$, if we assume a complete rejuvenation during shear melting ($\lambda_0 \approx 0$). The sufficiently aged system with $\lambda \gg 1$, can be represented as: $\tau \approx \tau_0 \tilde{t}^m$. We have already reported that $\tau$ follows a power law dependence on time with a power law coefficient $\mu$. Consequently, we have: $m \approx \mu$. On the other hand, Fig. 1 suggests that, after a sufficient time has elapsed, $G$ shows a power law dependence on time as $G \sim t^\nu$. It can be seen that $\nu \approx 0$ for the two hard sphere glasses and the Carbopol dispersion, while $\nu \approx 0.19$ for the 14 days old clay dispersion and $\nu \approx 0.12$ for the 1000 days old clay dispersion. This leads to $\eta \approx t^{\nu+\mu}$. In terms of time, Eq. (4) can be expressed as $\eta = \eta_0(1 + (\tilde{t} + \lambda_0)^n)$, and considering $\tilde{t} \gg \lambda_0$ as well as $\tilde{t} \gg 1$, we can approximate $n \approx \nu + \mu$. The values of $n$, $\nu$ and $\mu$ for all the five systems are listed in Table 1. Furthermore, differentiating the expression for the steady state flow curve given by Eq. (11) to obtain $d\tilde{\sigma}_{ss}/d\tilde{\dot{\gamma}}_{ss} = 0$ leads to estimation of $\lambda_c = (n-1)^{-1/n}$. Furthermore, assuming $\lambda_0 \approx 0$, the time at which $\lambda = \lambda_c$ is given by $\lambda_c = t_c/T_0$, while the critical stress is given by: $\tilde{\sigma}_c = \frac{1}{\beta}\frac{\eta_0}{T_0}\frac{(1+\lambda_c^n)}{\lambda_c}$. We use the experimentally obtained values of $\sigma_c$, $t_c$ and $\eta_0$ to obtain the remaining parameters $T_0$ and $\beta$.

**Table 1: Power law coefficient of material properties for Structure Kinetic Model obtained by fitting the experimental data**

| Parameters | 3.5 wt.% Clay dispersion | | Hard sphere glasses | | 0.2 wt.% Carbopol dispersion |
| --- | --- | --- | --- | --- | --- |
| | Rest Time | | $\Phi = 0.625$ | $\Phi = 0.595$ | |
| | 14 days | 1000 days | | | |
| $\mu$ | 2.35 | 1.13 | 0.9 | 0.95 | 0.97 |
| $n$ | 2.54 | 1.25 | 0.9 | 0.95 | 0.97 |
| $\nu$ | 0.19 | 0.12 | 0 | 0 | 0 |

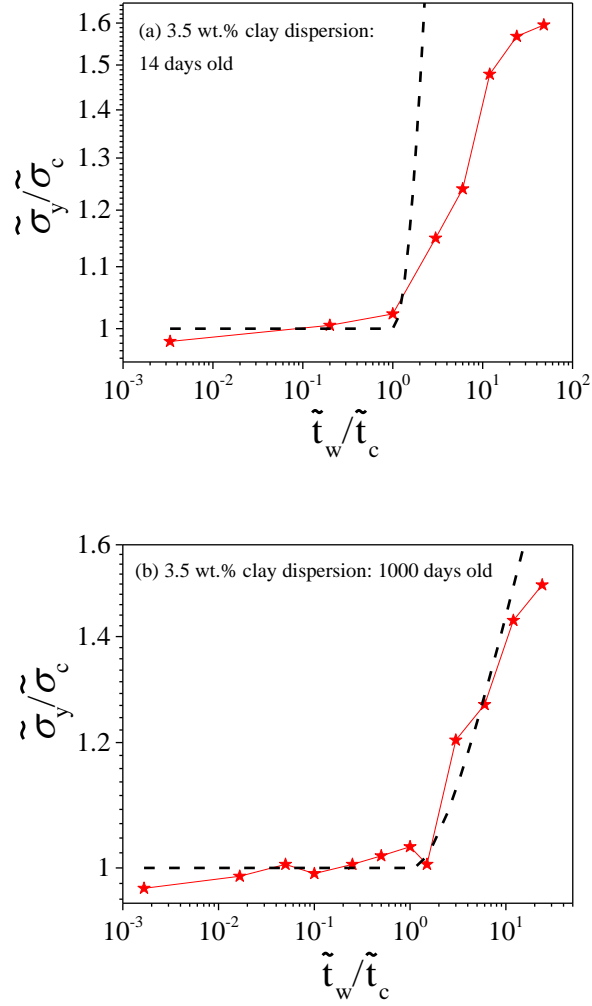

**Fig.9** Prediction of time dependent yield stress from the structure kinetic model with viscoelastic constitutive equation, as mentioned in Eq. (12) (dashed line) is plotted together with experimental data (red asterisk) for 3.5wt% clay dispersions of rest time (a) 14 days ($n = 2.54$, $T_0$=351.4 s, $\eta_0$=349 Pa.s, $\beta = 0.08$) and (b) 1000 days ($n = 1.25$, $T_0$=148.6 s, $\eta_0$=849.1 Pa.s, $\beta = 0.16$).

The prediction of Eq. (12) for the experimental data of the yield stress as a function of aging time is shown in Fig 9. It can be seen that for the 1000 days old clay dispersion the prediction of the model matches very well with the experimental time dependent yield stress. For the 14 days old clay dispersion, however, the slope of the time dependence of the yield stress predicted by the model is much steeper than that of the experimental data. According to Eq. (12), yield stress is expected to increase in a power law fashion with exponent $(n − 1)$, which for 14 days old clay dispersion is around 1.54.

However, the actual yield stress increases with much weaker dependence. A better prediction may require a more detailed structural kinetic model as well as the constitutive equation, for the clay system with 14 days of rest time. The model employed in the present work, on the other hand, is the simplest viscoelastic constitutive equation (Maxwell model) with a single relaxation mode. Nevertheless, considering the simplicity of the model, it does capture the phenomenology of the time dependent yield stress quite well. On the other hand, for the other three systems, $n$ is lower than 1 but close to unity, and hence the model predicts a monotonic steady state flow curve, with stress showing almost a plateau in the limit of small $\dot{\gamma}_{ss}$. Consequently, the model does not predict a time dependent yield stress as also observed in the experiments on the hard sphere glasses and Carbopol dispersion. Strictly speaking, the steady state behavior of the model – that shows $n = 1$ to be the threshold that separates the monotonic curve from the non-monotonic curve – depends upon various aspects of the model including the nature of time evolution equation for $\lambda$, the constitutive equation and how $\lambda$ relates with the viscosity and modulus (if any). As a result, the threshold that separates the monotonic curve from the non-monotonic one and the values of parameters for which the plateau in a limit of small $\dot{\gamma}_{ss}$ is detected is highly model specific [85]. Therefore, although the two studied hard sphere glasses and the Carbopol dispersion show $n$ to be lower than unity, the real steady state flow curve can very well show a stress plateau in a limit of low shear rate as all these systems do exhibit a yield stress, as shown in Figure 4(c), (d) and (e). What the present modelling exercise illustrates is that there is a direct relation between a time dependent yield stress and a non-monotonic flow curve within the framework of structural kinetic model. This also leads us to conclude that the thixotropic systems that do not show time dependent yield stress necessarily have a monotonic steady state flow curve.

**Concluding Remarks:**

In the present work, we investigate five experimental systems: two clay dispersions, two hard sphere glasses and a Carbopol dispersion. All five systems are thixotropic in nature according to the recent thixotropy criterion proposed by Agarwal and coworkers [17] that adheres to IUPAC definition of thixotropy [68]. In addition, the relaxation times of all these systems exhibit a power law dependence on waiting time elapsed since the cessation of shear rejuvenation ($t_w$) given by $\tau \sim t_w^\mu$ as represented by Eq. (10). We can divide these experimental systems into two categories, (i) those with constant yield stress and (ii) those with a time dependent yield stress. Interestingly the two systems that show a time dependent yield stress also show a noticeable increase of the elastic modulus with waiting time as well as a value of $\mu > 1$. On the other hand, the three systems that show constant yield stress do not show any significant variation of the elastic modulus as a function of time. In addition, these systems also exhibit a simple or sub-aging behavior with $\mu \leq 1$. Therefore, the present work unequivocally shows that constant yield stress may be a prominent feature of some thixotropic systems.

We also investigate a viscoelastic structural kinetic model. The results clearly show a direct correlation between the nature of the flow curve and the dependence of yield stress on time. We observe that a non-monotonic steady state flow curve in a structural kinetic formalism necessarily leads to a time dependent yield stress. The experiments on the two clay dispersions show that for a certain period of time after the cessation of shear rejuvenation, the yield stress remains constant before it subsequently starts increasing. Quite remarkably, the model prediction qualitatively matches this behavior very well. A non-monotonic steady state flow curve wherein the stress decreases with increase in shear rate is known to be an unstable region, where a steady homogeneous flow cannot be maintained [92]. Consequently, when the imposed shear rate is such that the flow curve has a negative slope, the flow field becomes inhomogeneous and gets banded into stable bands of different shear rates [99, 100]. Such shear banding in a thixotropic framework is a direct manifestation of the non-monotonic flow curve that also must get accompanied by a time dependent yield stress. However, it must be mentioned that not every kind of shear banding is due to a non-monotonic flow curve. It has been shown that similar hard sphere glasses as those studied here may show shear banding due to shear-concentration coupling along with the migration of particles to the low shear rate region. However, in this case, the flow curve does not show a non-monotonic shape, and the yield stress and $G'$ do not increase with waiting time [25, 101]. In general, the model suggests that the constant yield stress materials necessarily have a monotonic flow curve (with a plateau in a limit of low shear rates), which could be thixotropic. Consequently, thixotropy is not necessarily associated with only a non-monotonic flow curve.

The present work also shows that irrespective of whether the yield stress is constant or time dependent, a viscosity bifurcation is observed for all the investigated thixotropic materials. This suggests that the viscosity bifurcation is also not exclusively associated with non-monotonic constitutive relations in the thixotropic framework. Another phenomenon that is usually associated with thixotropy is the presence of rheological hysteresis. However, a recent report very clearly shows that non-linear viscoelastic (non-thixotropic) fluid shows various qualitative trends of hysteresis shown by thixotropic materials [102]. Therefore, hysteresis cannot be exclusively associated with thixotropy, irrespective of the nature of the flow curve followed by material, i.e., monotonic or non-monotonic.

We should also note that there could be materials that show thixotropic behavior without showing yield stress. For example, various small concentration colloidal dispersions (including clay dispersions), under quiescent conditions, undergo structure (aggregate) formation as a function of time, causing their viscosity to increase. Such structure (aggregates) can be broken by the application of a deformation field, leading to a time dependent decrease in viscosity. However, the aggregate formation could be so slow that the material does not show any yield stress over practically explorable timescales. Such materials, for all practical purposes, are thixotropic but without yield stress. Therefore, thixotropy does not necessarily imply the presence of yield stress (constant or time dependent), at least over practically explorable timescales.

Finally, it is important to understand the origin of thixotropy in a material. Any out of equilibrium material that is in a stress-free state undergoes microstructural evolution to lower its free energy. In structurally arrested materials such as colloidal gels, colloidal glasses, and various soft paste-like materials, etc., this process of evolution to lower the free energy is usually termed as physical aging. Physical aging necessarily causes an increase in relaxation time, leading to increase in viscosity. If the application of a deformation field reverses the process of physical aging, wherein relaxation time (and hence viscosity) decreases with time, the material is called thixotropic. However, the most fundamental ingredient of a material that facilitates physical aging is the thermal nature of its constituents that drive their microstructural evolution. If the constituents are athermal, the material will remain frozen forever in a high free energy state. Such material may show yield stress, but there will not be any time dependency in their rheological properties. We can therefore term such material, in its true sense, as simple (non-thixotropic) yield stress materials. Among the thixotropic materials, on the other hand, there could exist three possibilities, constant yield stress, time dependent yield stress and no yield stress. Furthermore, viscoelastic thixotropic materials can also be categorized based on whether the modulus remains constant and/or how the relaxation time evolves with time. However, irrespective of the categorization, it is important to understand that lowering of free energy under stress free conditions (that causes an increase in viscosity) and increase in free energy under application of deformation field (which results in a decrease in viscosity) to be the inherent aspect of thixotropy. The ensuing yield stress (if any) is a consequence of viscosity bifurcation which depends on whether for a given state of a material the applied stress results in a continuous increase in the viscosity or eventual attainment of constant steady state viscosity.

**Appendix:**

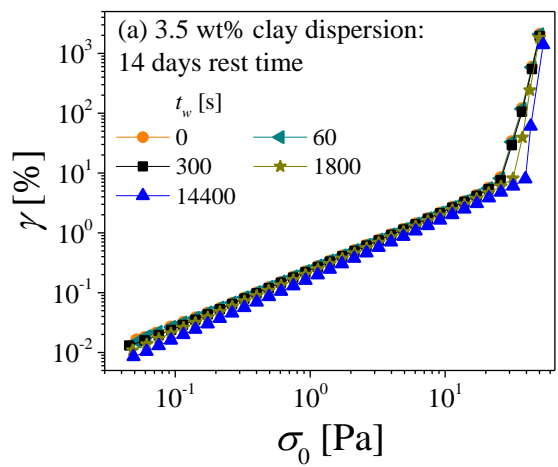

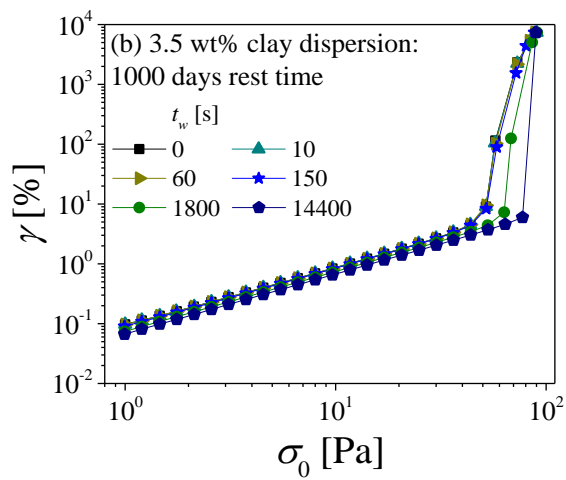

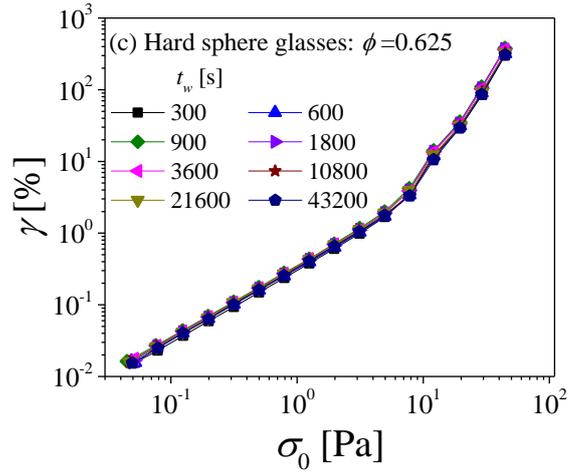

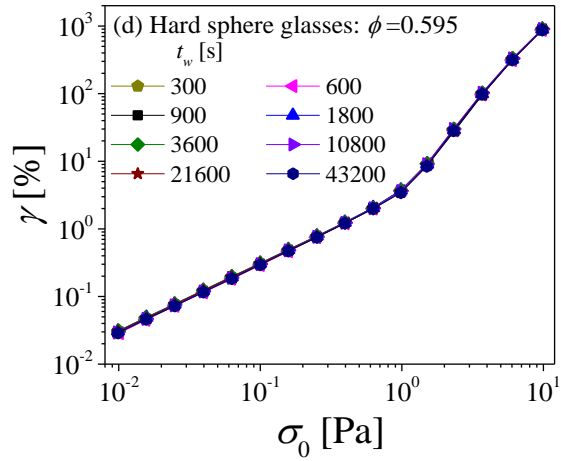

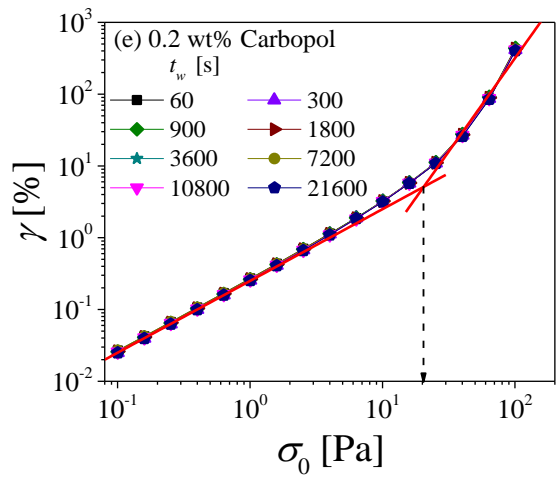

**Fig. A1** Evolution of $\gamma$ plotted as a function of increasing magnitude of oscillatory stress for 3.5 wt.% clay dispersion of rest time (a) 14 days and (b) 1000 days at $\omega$=0.63 rad/s; for hard sphere glasses with (c) $\phi$=0.625 and (d) $\phi$=0.595 at $\omega$=0.1 rad/s; for (e) 0.2wt% Carbopol at $\omega$=6.28 rad/s. The static yield stress, $\sigma_y$ is marked at the point of intersection

of the two slopes of the curve. The various symbols represent different $t_w$ as mentioned in the legends. The curves in (a) and (b) shows the time dependent static $\sigma_y$ observed in the explored range of $t_w$ for (a) 14 days ($\sigma_y \approx$ 25-40 Pa) and (b) 1000 days ($\sigma_y \approx$ 50-75 Pa) rest time of 3.5 wt.% clay dispersion. The curves in(c), (d) and (e) shows the time independent static $\sigma_y$ observed in the explored range of $t_w$ for hard sphere glasses with (c) $\phi$=0.625 ($\sigma_y \approx$ 8 Pa) and (d) $\phi$=0.595 ($\sigma_y \approx$ 1 Pa) and for (e) 0.2wt.% Carbopol ($\sigma_y \approx$ 20 Pa).

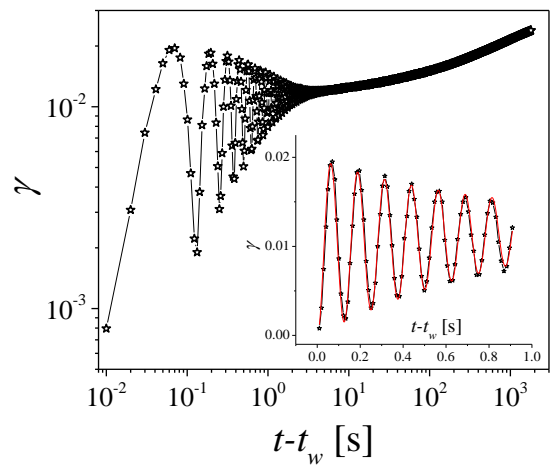

**Fig A2** A conventional time evolution curve of $\gamma$ shown for 3.5 wt% clay dispersion of 14 days rest time on application of step stress of $\sigma = 5$ Pa at aging time $t_w$=1800 s. The inset shows the fitting line for the initial damped oscillation given by Eq. (5) in creep data.

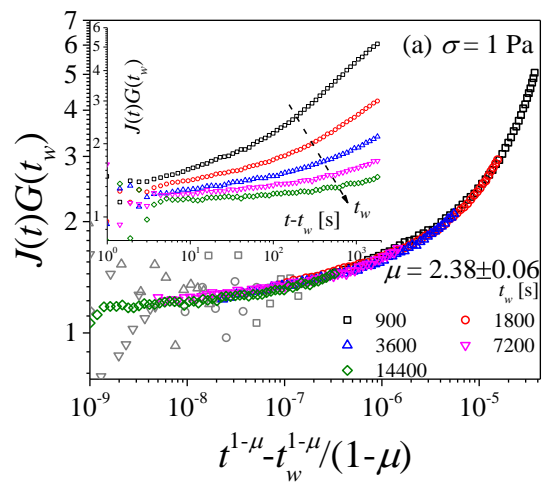

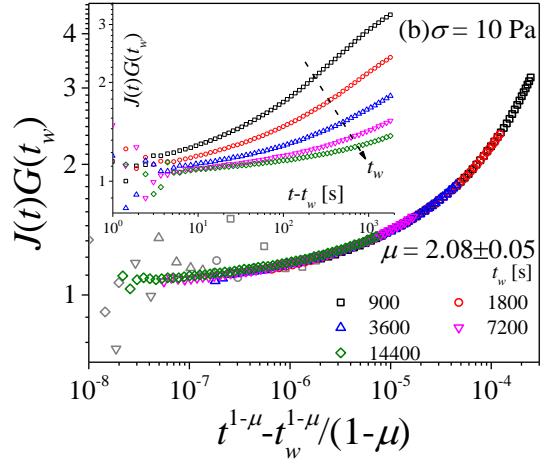

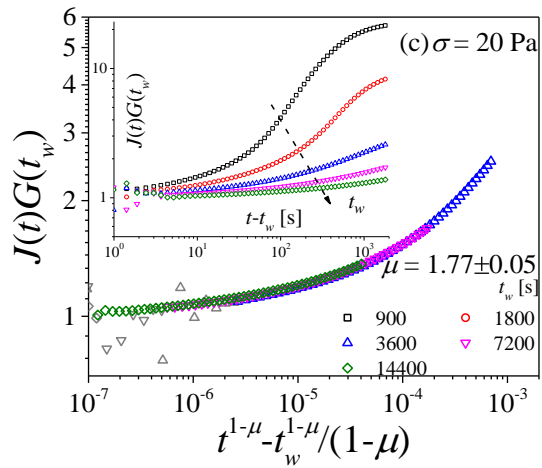

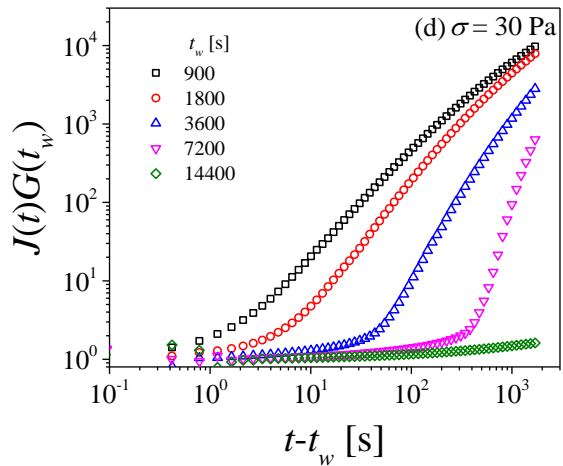

**Fig A3** Time aging time superposition of creep compliance obtained as a function of $t_w$ for a 3.5 wt.% clay dispersion with 14 days rest times at different stress magnitudes (a) $\sigma$ = 1 Pa ($\mu$ = 2.38), (b) $\sigma$ = 10 Pa ($\mu$ = 2.08), and (c) $\sigma$ = 20 Pa ($\mu$ = 1.77). In the insets the corresponding data for normalized compliance is shown as a function of creep time ($t - t_w$). The delayed yielding observed in creep compliance at higher creep time is shown in

(d) for $\sigma$= 30 Pa. The various symbols represent the different $t_w$ as mentioned in the legends. The method to obtain the vertical shift factor $G(t_w)$ is shown in supplementary information figure S2. The arrows are guide to the eyes in the increasing direction of $t_w$.

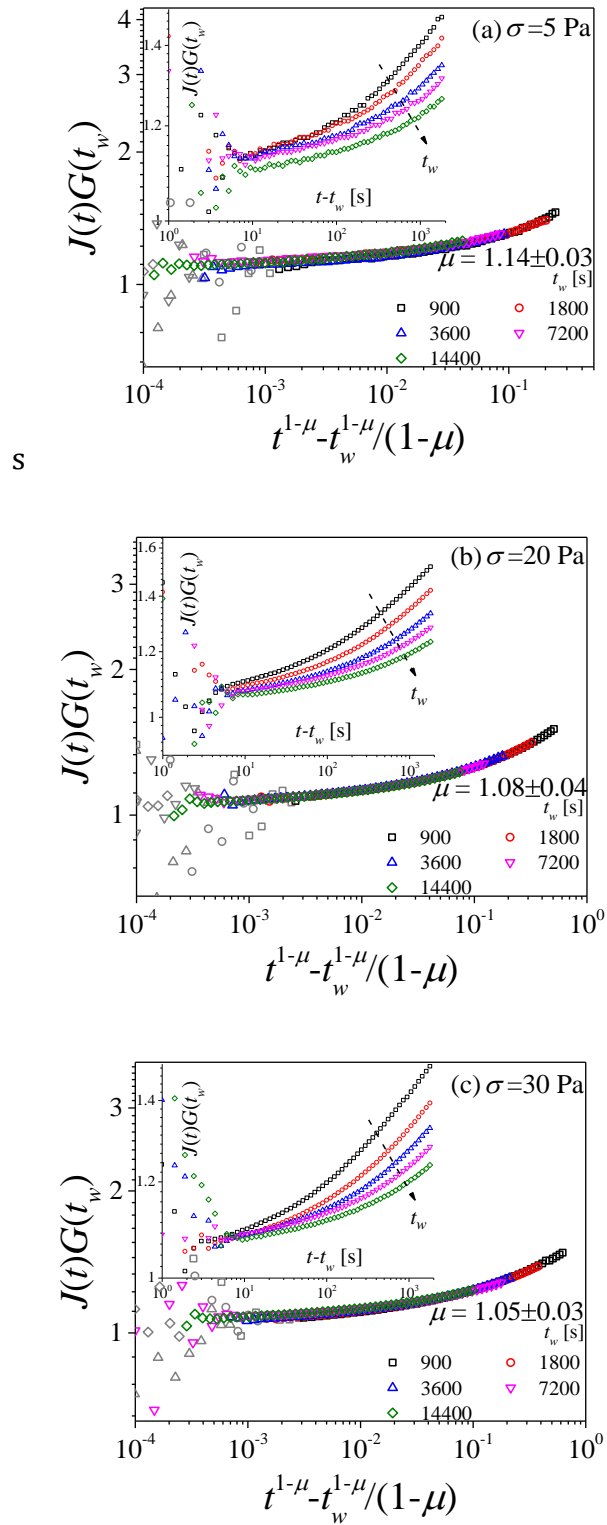

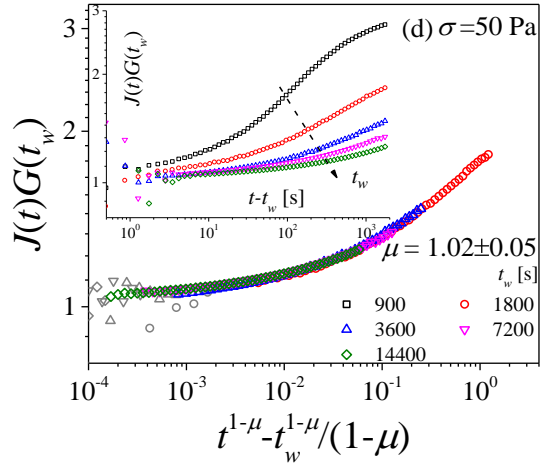

**Fig A4** Time aging time superposition of creep compliance obtained as a function of $t_w$ for a 3.5 wt.% clay suspension with 1000 days rest times at different stress magnitudes (a) $\sigma$= 5 Pa ($\mu$ = 1.14), (b) $\sigma$= 20 Pa ($\mu$ = 1.08), (c) $\sigma$= 30 Pa ($\mu$ = 1.05), and (d) $\sigma$= 50 Pa ($\mu$ = 1.02). In the insets the corresponding data for normalized compliance is shown as a function of creep time ($t - t_w$). The various symbols represent the different $t_w$ as mentioned in the legends. The method to obtain the vertical shift factor $G(t_w)$ is shown in supplementary information figure S2. The arrows are guide to the eyes in the increasing direction of $t_w$.

**Acknowledgement:** YMJ would like to acknowledge financial support from Science and Engineering Research Board, Government of India. GP acknowledges funding from EU by the European Soft Matter Infrastructure project EUSMI (grant agreement 731019) and the National Greek project Innovation-EL (MIS 5002772).

**Author Declarations:** The authors have no conflicts to disclose.